\documentclass[aps,pre,showpacs,reprint,superscriptaddress,author-year,amsmath,amsfonts]{revtex4-1}
\usepackage{graphicx}
\newcommand{\tlname}[1]{\ensuremath{\mathit{#1}}}

\usepackage[usenames,dvipsnames]{xcolor}
\usepackage[normalem]{ulem}
\let\ocite\cite
\def\cite#1{(\ocite{#1})}

\begin{document}
\title{Nonlinear rheology of dense colloidal systems with short-ranged
  attraction: A mode-coupling theory analysis}
\date\today
\newcommand\dlr{\affiliation{Institut f\"ur Materialphysik im Weltraum,
  Deutsches Zentrum f\"ur Luft- und Raumfahrt (DLR), 51170 K\"oln,
  Germany}}
\newcommand\ukn{\affiliation{Zukunftskolleg and
  Fachbereich Physik, Universit\"at Konstanz,
  78457 Konstanz, Germany}}
\author{Madhu~Priya}\dlr
\author{Thomas~Voigtmann}\dlr\ukn

\begin{abstract}
The nonlinear rheology of glass-forming colloidal suspensions with
short-ranged attractions is discussed within the integration-through
transients framework combined with the mode-coupling theory of the
glass transition (ITT-MCT). Calculations are based on the square-well
system (SWS), as a model for colloid-polymer mixtures. The
high-density regime featuring reentrant melting of the glass upon increasing
the attraction strength, and the crossover from repulsive glasses formed
at weak attraction to attractive glasses formed at strong attraction, are
discussed. Flow curves are found in qualitative agreement with experimental
data, featuring a strong increase in the yield stress, and, for suitable
interaction parameters, the crossover between two yield stresses.
The yield strain, defined as the position of the stress overshoot under
startup flow, is found to be proportional to the attraction range
for strong attraction. At weak and intermediate attraction strength,
the combined effects of hard-core caging and attraction-driven bonding result
in a richer dependence on the parameters.
The first normal-stress difference exhibits a weaker dependence on short-ranged
attractions as the shear stress, since the latter is more sensitive
the short-wavelength features of the static structure.
\end{abstract}

\maketitle

\section{Introduction}

The ability to fine-tune the rheological properties of colloidal suspensions
is of large importance in manufacturing, in particular of high-solid
dispersions \cite{Larson}.
A reduction in viscosity at constant solid loading helps
to flow these systems more efficiently;
this is typically achieved by adding smaller particles or controlling the
particle-size polydispersity (as first detailed in rheological context
by \textcite{Farris1968}), be it in ceramic processing
\cite{FunkDinger} or in food rheology \cite{Servais2002}.

Another way of adusting the viscosity of highly dense hard-particle
suspensions is to make use of effective attractive interactions among the
colloids \cite{Woutersen1991,Lewis2000,Pandey2012}.
In charge-stabilized suspensions of nearly
hard-sphere particles, such attractions can be fine-tuned by
adding nonadsorbing polymers to the solution. This has recently been described
by \textcite{Willenbacher2011} for technologically relevant
aqueous dispersions of polystyrene-microgel and latex particles.
Adding a few $\text{g}/\text{l}$ of a linear polyethylene-oxide
polymer resulted in a reduction in viscosity by up to two orders of
magnitude.

This procedure makes use of a depletion-attraction induced reentrant behavior
of the glass transition in the colloid-density--attraction-strength state
space. It is well known that the addition of free polymer to a hard-sphere
like colloidal suspension induces an entropic interaction among them: around
the colloid particles, a depletion layer deprived of polymer exists, but
on approach, two colloid particles can reduce the total volume of this
layer, resulting in an entropic gain for the distribution of the polymer
coils. If the polymer is treated as ideal-gas like, the resulting
effective interaction among the colloids is well described by the
Asakura-Oosawa potential \cite{Asakura1954}. For the purpose of a qualitative
discussion, a simpler model system is the square-well system (SWS),
where one simply assumes, relative to the hard-sphere diameter $d$,
a constant attraction of width $\delta$
(set by the radius of gyration of the free polymer) and
strength $U_0$ (controlled by the polymer concentration) surrounding the
hard-sphere core of the colloid. Together with the colloid packing fraction
$\varphi$, and normalizing energies by those of thermal fluctuations, $\Gamma
=U_0/k_BT$,
the three control parameters of the SWS thus are $(\varphi,\Gamma,\delta)$.

The SWS model at high density allows to distinguish repulsion-driven
glasses for small $\Gamma$, from attraction-driven glasses for large $\Gamma$.
This distinction was first established as a qualitative one
in the mode-coupling theory of the
glass transition (MCT) \cite{Fabbian1999,Bergenholtz1999,Dawson2001}, where for
small enough attraction range, it is signalled
by a glass--glass transition. Crossing this transition, the nonergodic
contribution to the average
structure of the amorphous solid changes abruptly, leading to a sharp
change in elastic moduli and related quantities.
The glass--glass transition results from a crossing of glass--transition lines
in the $\Gamma$--$\varphi$ plane, one leading to the ``repulsive glass''
(weakly dependent on $\Gamma$),
the other leading to the ``attractive glass'' (depending more strongly
on $\Gamma$ than on $\varphi$).
The repulsive glass is characterized by hard-core repulsion, and particles
are localized since they are trapped in ``cages'' formed by their neighbors.
As a result, individual-particle motion becomes localized at the
repulsive-glass transition on a length scale of typically around $10\%$
of a particle diameter. This is often referred to as the criterion of
Lindemann (who estimated a similar localization length close to the melting
point of crystals).
In the attractive glass on the other hand, particle localization is driven
by ``bonds'' formed between the particles, leading to a localization length
of the order of $\delta$. If $\delta\ll0.1$, the two arrest mechanisms
occur on separated length scales, leading to rich dynamical behavior in
the parameter regime where both transition lines meet. In particular,
MCT predicts a critical attraction range, $\delta^*\approx0.0465$
\cite{Dawson2001}, where a higher-order bifurcation singularity in the
glass-transition diagram appears. For $\delta<\delta^*$, the glass--glass
transition emerges, whose endpoint is also connected to a higher-order
singularity. These influence the dynamics in the vicinity: the well
documented two-step relaxation process of density fluctuations close to
the ordinary glass transition is replaced by leading-order asymptotic
logarithmic decay \cite{Goetze2002,Sperl2003}.

En route to the attractive glass, a weak attraction is found to destabilize
the repulsive glass, shifting its transition point to higher packing
fraction. This is thought to arise since the attraction-induced increase
of particle pairs in close contact inevitably opens up nearest-neighbor
cages \cite{Sciortino2002}.
As a result, one can find lines through the state space, say at
fixed packing fraction $\varphi$ and polymer size ratio $\delta$, where
a continuous increase in attraction strength first melts the repulsive
glass, and then re-freezes the resulting liquid as an attractive glass.
This ``reentry phenomenon'' is the basis of the viscosity reduction
discussed by \textcite{Willenbacher2011}.

The predictions of MCT for the quiescent dynamics of the
high-density SWS have been extensively tested 
in simulation \cite{Sciortino2003} and experiment \cite{Eckert2002,Pham2002,Pham2004}.
The theory gives a good qualitative account of the phenomena there,
although especially in the attractive glass, certain hopping-like
relaxation modes that destroy bonding-induced particle localization are
not accounted for in MCT.

Since the attractive glass lines extends to low $\varphi$, it is tempting
to connect this low-density attractive-glass transition predicted by MCT with
the gel transition, and several studies address the low-density limit \cite{Bergenholtz1999,Bergenholtz1999b,Bergenholtz2000,Shah2003,Kroy2004}.
However, gelation involves spatial heterogeneities
outside the scope of MCT, so a subtle distinction between (low-density) gels
and (high-density) attractive glasses has to be maintained
\cite{Puertas2002,Puertas2003,Puertas2004,Zaccarelli2009,Laurati2009,Laurati2011,Koumakis2011}.
Numerous of rheology studies have discussed gelation-induced phenomena
in detail (see, e.g., \textcite{Rueb1998,Prasad2003,Lindstrom2012} and
references therein).
Only more recently, the high-density attractive-glass regime has caught
attention.
We restrict ourselves to this regime in the following.

The rheological signature of the reentrant glass transition in the SWS, but
also in related binary mixtures of particles with very different sizes,
in terms of linear-response shear moduli has been discussed before
\cite{Grandjean2004,Sztucki2006,Narayanan2006,Mayer2007,Truzzolillo2013}.
Since the relevant length scale
changes, one notes a strong increase with increasing attraction, from
values $\sim k_BT/d^3$ (where $d$ is the colloid diameter), to values
$\sim k_BT/d^3/\delta$ (if the attraction strength is kept constant)
\cite{Bergenholtz2000}.

An intriguing phenomenon arising in the nonlinear rheology of attractive
glasses was first pointed out in seminal papers by
\textcite{Pham2006,Pham2008}. Using strain sweeps of large-amplitude oscillatory
rheology, they found a two-step yielding, indicated by two distinct peaks
in the stress--strain-amplitude curves: the first emerges at the end of
the reversible (almost) linear response regime at around $\gamma_y\approx10\%$
of strain amplitude, while the second appears at much higher amplitudes,
($50\%$ to $100\%$ in their study).
The first peak is reminiscent of the yielding of hard-sphere repulsive
glasses: there, stress--strain curves typically show a nonmonotonic
crossover between the (almost) linear elastic-like increase at low strain to
the steady-state stress approached in the flowing state. This ``stress
overshoot'' \cite{Zausch2008,Koumakis2012,Amann2013} is interpreted as the point where
nearest-neighbor cages first break, in agreement with its position $\gamma_y$
and the Lindemann criterion. One indeed observes that the first yielding
step occurs at lower $\gamma_y$ for lower $\delta$ \cite{Koumakis2011}.
The second peak observed in attractive glasses
is now generally interpreted as the breakup of larger structural units (such
as clusters) \cite{Laurati2011}
that can still form even if individual bonds are broken.

Theoretical descriptions of the yielding behavior of attractive glasses
are still scarce. Notably, \textcite{Kobelev2005,Kobelev2005b}
were the first to address the limit of attractions strong compared to thermal
fluctuations, using the notion of barrier crossing in a nonequilibrium
free-energy landscape that is modified by shear.

MCT has been extended to describe the nonlinear rheology of shear-thinning
colloidal suspensions by Fuchs and Cates \cite{Fuchs2002c,Fuchs2009,Brader2010},
in a framework called integration through transients (ITT).
This theory starts from the nonequilibrium Smoluchowski equation and
contains coupling coefficients that are given in terms of the equilibrium
static structure functions. Its application to the SWS is hence
straightforward, but lacking so far. In this contribution, we employ
the ITT-MCT with an additional isotropic assumption for its memory function,
to the SWS model studied previously in the quiescent state. We specifically
address the flow curves in the vicinity of the glass--glass transition,
and discuss the dependence of the ITT-MCT yield strain on attraction range
$\delta$ and strength $\Gamma$.

The paper is organized as follows: in Sec.~\ref{theory} we outline the
theory together with technical details of the calculation. In Sec.~\ref{results}
we present results for flowcurves and stress--strain relations, followed
by a discussion of the results in light of experimental findings
(Sec.~\ref{discussion}).
Section~\ref{conclusion} contains conclusions.

\section{Theory}\label{theory}

The ITT-MCT description of nonlinear rheology of glass-forming colloidal suspensions
is based around transient correlation functions, in particular those
of colloid number-density fluctuations of the $N$-particle system,
$\varrho_{\vec q}=\sum_{k=1}^N\exp[i\vec q\cdot\vec r_k]$.
In the case of an imposed time-independent shear rate,
the transient density correlator
$\phi_{\vec q}(t)=\langle\varrho_{\vec q}^*\exp[\Omega^\dagger t]
\varrho_{\vec q(t)}\rangle/\mathcal N$ measures the overlap of a density
fluctuation at wave-vector $\vec q$ with one at a time $t$ earlier whose
wavevector $\vec q(t)$ evolves due to flow-induced advection to become $\vec q$.
Here, $\Omega^\dagger$ is the adjoint Smoluchowski operator describing
the time evolution of dynamical variables of the colloidal system, i.e.,
the Brownian motion of $N$ interacting particles on top of the prescribed flow
field. Hydrodynamic interactions are neglected in the theory.
The angular brackets denote the canonical equilibrium average, and
$\mathcal N$ is a normalization factor to set $\phi_{\vec q}(0)=1$.

In principle, the wave-vector dependent correlation functions under flow
are anisotropic. However, structural anisotropies of sheared dense colloidal
suspensions are found to be rather small \cite{Henrich2009,Koumakis2012},
at least at small bare P\'eclet numbers, $\dot\gamma\tau_0\ll1$, with the
Brownian relaxation time of a single particle $\tau_0$.
One can hence consider an ad-hoc simplification of ITT-MCT where the
wave-vector dependence of all correlation functions is assumed to be
isotropic. This approximation was first applied to discuss the qualitative
features of the hard-sphere system under shear, where it was termed
the isotropically sheared hard-sphere model (ISHSM) \cite{Fuchs2003,Fuchs2009}.
The full anisotropic ITT-MCT equations are numerically rather demanding and
have been solved so far only in two spatial dimensions \cite{Henrich2009}.
Only very recently, the full 3D theory has been solved.
This scheme has now also been applied to the SWS model \cite{AmannJOR},
but only a selected number of state points for one particular attraction
range could be studied. The attraction-induced sensitivity of the MCT integrals
to a much larger wave-vector range as for hard spheres \cite{Dawson2001}
makes it desirable to employ large numerical grids in the numerical evaluation
of the theory. This is something that is not yet easily done in the full
3D anisotropic calculation.
The present study hence employs the isotropic approximation,
in order to provide a qualitative account of a wider parameter range.
Comparing with \textcite{AmannJOR}, the effects of the isotropization
of the theory can be checked; they are found to be quantitative, but
not qualitative.

The isotropic ITT-MCT equation of motion for the transient density correlation
function in steady shear is given by
\begin{equation}\label{phit}
  \dot\phi_q(t)+\Gamma_q\left\{\phi_q(t)
  +\int_0^tdt'\,m_q(t-t')\dot\phi_q(t')\right\}=0\,.
\end{equation}
Here, $m_q(t)$ is the memory kernel describing the combined effects of
slow structural relaxation and its modification through shear. In the
isotropic model one sets
\begin{equation}\label{mt}
  m_q(t)\approx\frac1{2V}S_q\sum_{\vec k}V^{(\dot\gamma)}_{\vec q,\vec k}(t)
  S_kS_p\phi_k(t)\phi_{|\vec q-\vec k|}(t)\,,
\end{equation}
with the mode-coupling vertex that is time dependent due to the advection
in shear flow. This vertex is, as usual, determined by the equilibrium,
quiescent direct correlation
functions (DCF) $c_k$ alone. The DCF is connected to the static structure
factor of the sytstem, $S_k=1/(1-nc_k)$, where $n$ is the colloid number
density. One has
\begin{equation}\label{vqk}
  V^{(\dot\gamma)}_{\vec q,\vec k}(t)=
  \frac{n}{q^4}\left[\vec q\cdot\vec kc_{k(t)}+\vec q\cdot\vec p
  c_{p(t)}\right]\left[\vec q\cdot\vec kc_k+\vec q\cdot\vec pc_p\right]
  \,.
\end{equation}
Here, $k(t)=k\sqrt{1+(\dot\gamma t/\gamma_c)^2/3}$ is the advected wave vector
in the isotropic approximation. The dimensionless model parameter $\gamma_c$
is included here, following a suggestion originally discussed in the
schematic-model simplification of ITT-MCT \cite{Brader2009}, to adjust
the scale of the repulsive-glass cage breaking. We fix this parameter
to $\gamma_c=0.1$, independent on density, attraction strength or width.

In the quiescent state, MCT predicts the long-time behavior of $\phi_q(t)$
to change at certain critical control parameters, identified as idealized
glass transitions \cite{Goetze}. For small enough coupling
strength in the memory kernel, $\phi_q(t)$ decays to zero as expected for
a liquid. For stronger coupling, a finite nonergodicity parameter
(also called glass form factor), $f_q=\lim_{t\to\infty}\phi_q(t)>0$, arises
and signals dynamical arrest of density fluctuations.
The $f_q$ are determined by a set of nonlinear equations, and the
transition points of MCT are bifurcation points of these equations.
These ideal glass transitions can be classified into the simplest sequence of bifurcation transitions,
termed $\mathcal A_\ell$ bifurcations with $\ell=2,3,\ldots$ in the mathematical
literature.
The main contributions to the wave-vector dependent coupling of density
correlators close to such $\mathcal A_\ell$ transitions stem from intermediate
$q$.
Wave-vector advection under shear decorrelates the contributions from the
DCF appearing in the MCT vertices, Eq.~\eqref{vqk}, and hence counteracts
dynamical arrest. This is the basis of shear thinning in ITT-MCT.
It also implies that under steady shear, any finite shear rate will
melt the glass \cite{Fuchs2002c}.

To derive a nonlinear constitutive equation for the non-Newtonian stress
tensor based on ITT-MCT, one starts from the full anisotropic nonlinear
generalization of a Green-Kubo relation
\cite{Fuchs2002c,Brader2007,Brader2008}. In the isotropic approximation,
one writes
\begin{equation}\label{sigma}
  \boldsymbol\sigma(t)
  =\int_0^tdt'\,\left[-\frac{\partial}{\partial t'}\boldsymbol B(t-t')\right]
  G(t-t')\,,
\end{equation}
assuming time-independent shear to start at $t=0$ from a stress-free
equilibrium state.
Here, the Finger tensor (or left Cauchy-Green tensor) appears,
\begin{equation}
  \boldsymbol B(t-t')=\begin{pmatrix}1+\gamma_{tt'}^2 & \gamma_{tt'} & 0\\
  \gamma_{tt'}&1&0\\ 0&0&1\end{pmatrix}\,,
\end{equation}
for simple shear with a velocity gradient $\dot\gamma=\partial_yv_x$.
The accumulated strain is given by $\gamma_{tt'}=\int_{t'}^t\dot\gamma(s)\,ds$,
or simply $\dot\gamma(t-t')$ for steady shear.
The Finger tensor ensures that Eq.~\eqref{sigma} obeys the principle
of material objectivity \cite{Salencon}.

For the nonlinear dynamic shear modulus $G(t)$, different isotropic ITT-MCT
approximations exist \cite{Fuchs2009}. In the following, we use the
original proposition \cite{Fuchs2003},
\begin{equation}\label{gt}
  G(t)=\frac{k_BT}{60\pi^2}\int dk\,k^4\frac{S'_kS'_{k(t)}}{S^2_{k(t)}}
  \phi_k^2(t)\,,
\end{equation}
which we find to give most consistent results overall.
From Eqs.~\eqref{sigma} and \eqref{gt}, one obtains the shear stress
as
\begin{equation}\label{sigxy}
  \sigma_{xy}(t)=\dot\gamma\int_0^tdt'\,G(t')\,.
\end{equation}
We will denote the steady-state value of the shear stress by
$\sigma=\lim_{t\to\infty}\sigma_{xy}(t)$. It is used to define the shear
viscosity in the usual way, $\eta=\sigma/\dot\gamma$.

Equation~\eqref{sigma} also allows to calculate normal-stress differences,
$N_1=\sigma_{xx}-\sigma_{yy}$ and $N_2=\sigma_{yy}-\sigma_{zz}$.
In the full anisotropic ITT-MCT, both a first and a second normal-stress
difference appear \cite{Farage2013,AmannJOR}, but typically the latter
is an order of magnitude smaller. In the isotropic model used here,
the second normal-stress difference is identically zero, $N_2=0$.
The first normal-stress difference is given by
\begin{equation}\label{n1}
  N_1=2\dot\gamma^2\int_0^\infty dt\,t\,G(t)\,,
\end{equation}
yielding the first normal-stress coefficient, $\Psi_1=N_1/\dot\gamma^2$.

It is apparent from above that both the viscosity and the
first normal-stress coefficient are given through very similar wave-number
integrals, only differing by an additional factor $t$ in the time integral
for $\Psi_1$. We define the corresponding wave-number dependent integrands
$I_\eta$ and $I_{\Psi_1}$ to discuss the relevance of different contributions
coming from different wave lengths, through $\eta=\int I_\eta(k)\,dk$
and $\Psi_1=\int I_{\Psi_1}(k)\,dk$, following a similar definition proposed
by \textcite{Farage2013}.
Explicitly, in the isotropic ITT-MCT model,
\begin{equation}\label{ieta}
  I_\eta(k)=\int_0^\infty dt\,
  \frac{k_BT}{60\pi^2}k^4\frac{S'_kS'_{k(t)}}{S^2_{k(t)}}\phi_k^2(t)\,,
\end{equation}
and
\begin{equation}\label{ipsi}
  I_{\Psi_1}(k)=\int_0^\infty dt\,t\,
  \frac{k_BT}{60\pi^2}k^4\frac{S'_kS'_{k(t)}}{S^2_{k(t)}}\phi_k^2(t)\,.
\end{equation}
The time dependence of the advected wave-vectors cause these expressions
to display different features as function of $k$, as we will discuss further
below.

Equations \eqref{phit} to \eqref{vqk} are solved numerically using
a standard procedure adapted to the calcuation of slowly decaying correlation
functions \cite{Fuchs2003}. From $\phi_q(t)$, one obtains $G(t)$ and hence
$\sigma(t)$ as well as $N_1(t)$ by a simple integration using the trapezoidal
rule. In solving the ITT-MCT equations numerically, one has to specify
a discrete grid of wave vectors, including both a low-$q$ and high-$q$
cutoff. The latter has to be chosen large enough, since the DCF entering
the MCT vertex will exhibit a slowly decaying power-law contribution,
$c_q\sim\sin[qd(1+\delta/2)]/q$ in a range $\pi\ll qd\ll\pi/\delta$,
before crossing over to its true $1/q^2$ asymptote \cite{Dawson2001}.
. Hence for small $\delta$, the large-$q$
structure of the vertex is decisive.
We fix the upper wave-number cutoff to $Q=400/d$ for the calculation of
the dynamics under shear. At this value, the DCF
has decayed to zero to within less than $0.5\%$ for all the state points
considered below. An equidistant grid with step size $\Delta q=0.2/d$ is
chosen. Calculations for the quiescent glass-transition points have been
obtained using $Qd=240$ and $\Delta q\,d=0.4$.
The unit of length is fixed by the diameter of the hard spheres,
$d=1$, and $k_BT$ is the unit of energy. The Brownian unit of time for a
colloidal
system is then given by fixing the single-particle free diffusion coefficient,
$D_0$, as $\tau_0=d^2/D_0=1$.

The quiescent static structure of the SWS model is evaluated in the
mean-spherical approximation (MSA) and leading order of small well width
$\delta$. Here, an analytic expression for the DCF is readily available
\cite{Dawson2001}, allowing to efficiently calculate the MCT vertex.

\section{Results}\label{results}

\begin{figure}
\includegraphics[width=.9\linewidth]{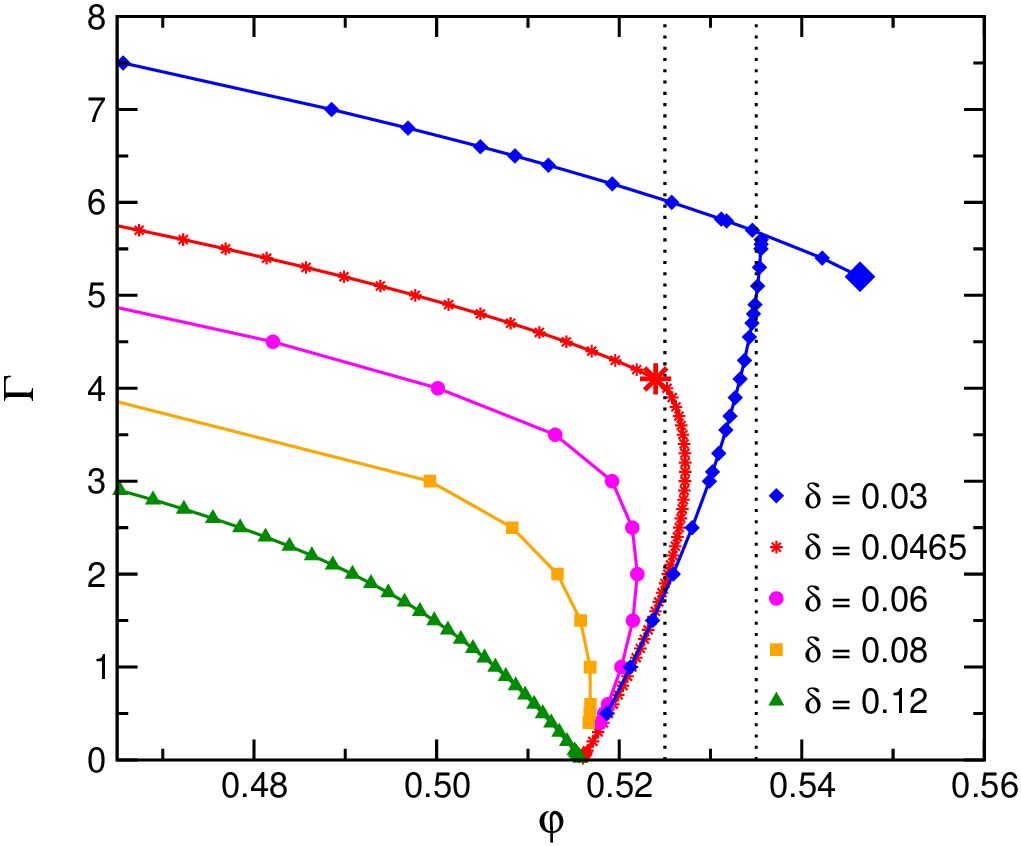}
\caption{\label{gt-diagram}
  Quiescent glass-transition diagram of the square-well system. Transition
  lines are shown in the packing-fraction--attraction-strength plane
  $(\varphi,\Gamma)$, for various attraction ranges $\delta$ as indicated.
  Dotted lines indicate paths considered in the following figures.
  For the $\delta=0.0465$ and $\delta=0.03$ curves, the bigger symbols mark
  approximate locations of the $\mathcal A_4$ and $\mathcal A_3$ higher-order
  singularities.
}
\end{figure}

Figure~\ref{gt-diagram} shows the glass-transition diagram of the
square-well system calculated from quiescent MCT \cite{Dawson2001}.
The $\Gamma=0$ line reflects the pure hard-sphere system. Here, MCT
predicts a glass transition at a critical packing fraction
$\varphi_c\approx0.516$. At large attraction range, $\delta>\delta^*$,
this transition point extends into the $\varphi$--$\Gamma$-plane as a
continuous line of glass transition points. According to MCT, these are
bifurcations in the long-time behavior of the density correlation function,
$f_q=\lim_{t\to\infty}\phi_q(t)$, where two solution branches of the nonlinear
equations determining the nonergodicity factors $f_q$ meet. They are termed
$\mathcal A_2$ singularities in the mathematical classification of bifurcations.

Lowering $\delta$ below a critical value $\delta^*\approx0.0465$, one
recognizes at fixed attraction range two transition lines:
a repulsive- and an attractive-glass line cross at an
intersection point. At this point, the repulsive-glass transition line
(extending from the hard-sphere reference point)
terminates, while the attractive-glass transition line continues as a
glass--glass transition, terminating in an endpoint.
Since all regular (repulsive and attractive) transition points are
$\mathcal A_2$ singularities, this endpoint marks a higher-order
singularity of type $\mathcal A_3$. At $\delta=\delta^*$, this endpoint
coincides with the crossing point, resulting in an even higher-order
singularity, of type $\mathcal A_4$.
The occurence of singularities of the $\mathcal A_\ell$ type is a
topologically robust feature: changing the interaction potential
somewhat, or modifying the approximations used in calculating the
static structure functions, will not change the scenario qualitatively.

One also recognizes in Fig.~\ref{gt-diagram} a reentry phenomenon when
increasing the attraction strength $\Gamma$ starting from the hard-sphere
case, $\Gamma=0$. For sufficiently short-ranged attractions, the
repulsive-glass transition line first shifts to
higher density, widening the region of the fluid and destabilizing the glass.
Hence there exists a narrow density window, where the hard-sphere system
is an ideal glass according to MCT, and first melts and then re-freezes
as the attraction strength is increased. In agreement with the physical
picture mentioned in the introduction \cite{Sciortino2002}, this only
occurs if the attraction range exceeds the hard-core-caging length scale
estimated by Lindemann, $\delta\gtrsim0.1$. Intuitively, adding an
attraction longer than Lindemann's localization length will not further
bond particles at high density.

\begin{figure}
\includegraphics[width=.9\linewidth]{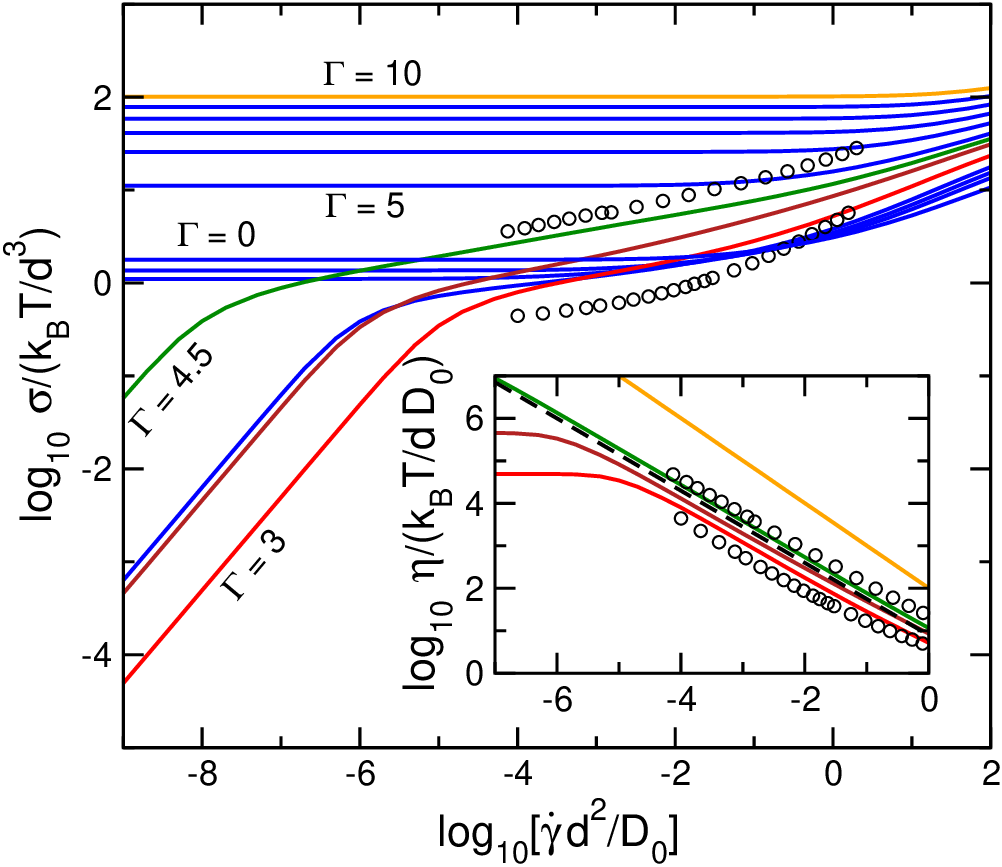}
\caption{\label{d04flowcurve}
  Flowcurve $\sigma(\dot\gamma)$ for the steady-state shear stress as a
  function of shear rate, for the square-well system with $\delta=0.04$,
  at fixed packing fraction $\varphi=0.525$, for increasing attraction
  strength (lines from bottom to top at high $\dot\gamma$:
  $\Gamma=0$, $1.0$, $1.5$, $2.0$,
  $3.0$, $4.0$, $4.5$, $5.0$, $6.0$, $7.0$, $8.0$, $9.0$, and $10.0$.
  Inset: shear viscosity $\eta=\sigma/\dot\gamma$ corresponding to the
  $\Gamma=3.0$, $4.0$, $4.5$, and $10.0$
  curves shown in the main figure. The dashed line indicates a power law
  $\propto\dot\gamma^{-0.85}$.
  Symbols are experimental data from \textcite{Pham2008}, converted
  setting $k_BT/d^3\equiv3.5\,\text{Pa}$ and $\tau_0=d^2/D_0\equiv0.002\,
  \text{s}$.
}
\end{figure}

We begin by discussing typical flow curves upon crossing from the repulsive
into the attractive glass. Figure~\ref{d04flowcurve} shows the case
$\delta=0.04$ at fixed density $\varphi=0.525$, with increasing attraction
strength $\Gamma$. At $\Gamma=0$, the system is a repulsive glass, and
exhibits a flowcurve of Herschel-Bulkley type: for $\dot\gamma\to0$, a
finite dynamic yield stress, $\sigma_y=\lim_{\dot\gamma\to0}\sigma(\dot\gamma)>0$,
characterizes the flow-melted glass.
The yield stress arises, because according to ITT-MCT, the nonlinear dynamic
shear modulus $G(t)$ decays on the flow-induced time scale
$\tau_{\dot\gamma}\sim1/\dot\gamma$; in the glass, this is the only relevant
time scale for long-time structural relaxation. In this regime, achieved for
$\dot\gamma\to0$, $G(t)$ becomes a function of $\dot\gamma t$ only, and
the integral determining $\sigma$, Eq.~\eqref{sigxy}, becomes independent
on $\dot\gamma$ \cite{Fuchs2002c}.
The $\sigma$-versus-$\dot\gamma$ curve monotonically increases with increasing
$\dot\gamma$. The limit of very high shear rates, $\dot\gamma\tau_0\gg1$,
will not be discussed further in the following. It probes the interplay
of flow and the short-time dynamics of the colloidal suspension, and MCT
is not designed to treat this properly.
The yield stress of the repulsive glass, $\sigma_y^\text{rep}$,
is roughly given by $k_BT/d^3$, the entropic energy scale of the system.

As $\Gamma$ is increased, the yield stress decreases slightly, indicating
a weakening of the glass structure with weak attraction. Increasing
$\Gamma$ further, the reentrant glass transition is probed: as the
quiescent glass state is molten by attraction, the flow curves show a
Newtonian regime, $\sigma\propto\dot\gamma$, at low shear rates, indicative
of a constant linear-response low-shear viscosity. The linear-response
regime extends over small dressed P\'eclet numbers, $\tlname{Pe}=\dot\gamma
\tau\ll1$, where $\tau$ is the structural relaxation time of the fluid.
For $\tlname{Pe}\gg1$, a sublinear increase in $\sigma(\dot\gamma)$ is
reminiscent of the yield-stress plateau in the glass.

Further increasing $\Gamma$, the Newtonian low-shear viscosity first
decreases, and then increases again, which is seen in the flow curves as
first the lowering, and then the increase in the linear regime. Finally,
the attractive glass is entered, and the flowcurves (in the ideal glass of
ITT-MCT) display a true yield
stress again. This yield stress of the attractive glass,
$\sigma_y^\text{attr}$, is considerably higher than its repulsive
counterpart. For the case shown in Fig.~\ref{d04flowcurve}, it starts
at approximately $10\,k_BT/d^3$ for $\Gamma=5$, and further increases
to around $100\,k_BT/d^3$ for $\Gamma=10$. This strong increase is also
responsible for the rise of the flow curves in the reentrant-fluid regime
at intermediate $\dot\gamma$, where it leads to a crossing of the
different $\sigma$-versus-$\dot\gamma$ curves for different $\Gamma$.

The yield stress increases by roughly a factor of ten when first crossing from
the repulsive into the attractive glass. Although the numerical value will
depend on the packing fraction, particle-size polydispersity and the
details of the interaction potential, our result is in good agreement with
the experimental data of \textcite{Pham2008}. Flowcurves
measured by these authors, on a suspension of PMMA hard-sphere-like colloids
with comparable attraction range, are shown in Fig.~\ref{d04flowcurve}
as symbols. In this comparison, only two unknown scale factors are fixed:
shear rates $\dot\gamma$ given in absolute time units have to be converted
to bare P\'eclet numbers $\dot\gamma\tau_0$
assuming a value for $\tau_0=d^2/D_0$.
From the Stokes-Einstein expression, one would obtain $\tau_0\approx
0.01\,\text{s}$
for the system of \textcite{Pham2008}; however, taking into account the
slowing down of free diffusion at high densities due to hydrodynamic
interactions, we allow this value to be adjusted, setting
$\tau_0=0.002\,\text{s}$.
The second parameter concerns the conversion of stresses into natural units.
Given the approximations involved in our calculation for the
simple SWS model, the agreement is reasonable.
An even stronger increase in yield stress, by about a factor $100$, has
been found by \textcite{Pandey2012} upon increasing the
attraction strength even further; this is again in quantitative agreement
with our Fig.~\ref{d04flowcurve}.

The inset of Fig.~\ref{d04flowcurve} displays the shear
viscosities corresponding to selected flow curves shown in the main panel.
The ideal-glass yield stress corresponds to shear-thinning
behavior with a trivial shear-thinning exponent $x=1$, i.e.,
$\eta\sim1/\dot\gamma$. This regime is barely reached for the lowest shear rates
in the experimental repulsive-glass data. The attractive-glass data do not
show the approach to this asymptote within the experimentally accessible window.
In the range covered there, an effective power law, $\eta\sim1/\dot\gamma^x$
with $x<1$ may be fitted to the data. This is apparent for the
$\Gamma=4.5$ curve in our ITT-MCT results, where $x\approx0.85$ may
reasonably be fitted to the $\eta$-versus-$\dot\gamma$ curve over more
than four orders of magnitude of variation in shear rate.
Such non-trivial shear-thinning exponents are routinely used to describe
experimental data \cite{Larson,Brader2010}, and have also been
emphasized by \textcite{Kobelev2005b}. We stress that
within ITT-MCT, they only appear as effective power laws describing the
crossover from repulsive to attractive glass; both glass types are, within
this theory, ultimately characterized by shear thinning with $x=1$ at
shear rates approaching zero.

The nonmonotonic variation in flow curves caused by the reentrant glass
transition also translates to the viscosities. However, only the
low-$\tlname{Pe}$ regime of the curves reflects the quiescent ideal
glass transition (defined as the point where the low-shear MCT viscosity
diverges). Since the $\sigma$-versus-$\dot\gamma$ curves for different
intermediate $\Gamma$ cross, the viscosities probed at finite shear rates
may exhibit a different ordering with respect to attraction strength.

\begin{figure}
\includegraphics[width=.9\linewidth]{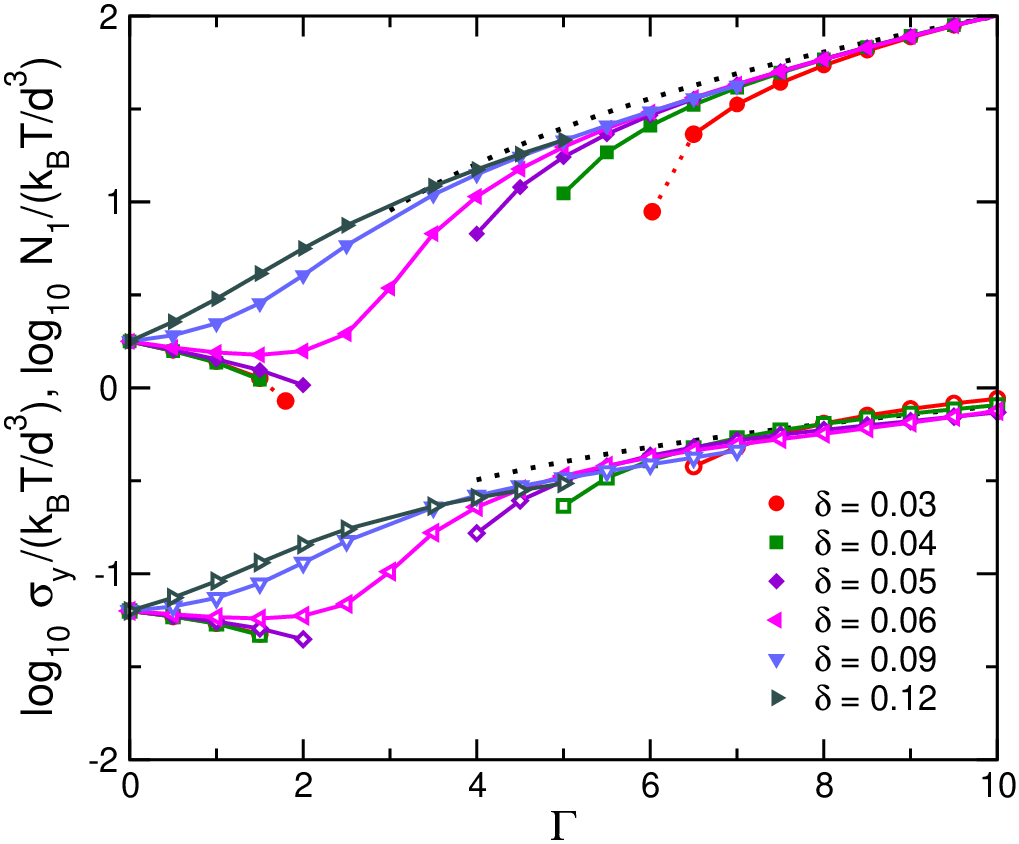}
\caption{\label{syGamma}
  Dynamic yield stress $\sigma_y=\lim_{\dot\gamma\to0}\sigma(\dot\gamma)$,
  as a function of attraction strength $\Gamma$ at fixed packing fraction
  $\varphi=0.525$, for various ranges $\delta$ as labeled
  (filled symbols). Data has been obtained setting $\dot\gamma=10^{-9}$,
  showing only points where this shear rate is in the asymptotic regime.
  For the $\delta=0.03$ curves, the points joined with dotted lines indicate
  the approach to the glass-transition points along the chosen path.
  Open symbols are the corresponding normal-stress differences $N_1$.
  Dotted lines indicate the relations $\sigma_y\propto\Gamma^2$ (upper) and
  $N_1\propto\Gamma$ (lower).
}
\end{figure}

The dependence of the yield stress on the attraction parameters is exemplified
in Fig.~\ref{syGamma} for those attraction strengths where the system
remains glassy at the chosen fixed packing fraction. As anticipated from
above, the repulsive-glass branch of $\sigma_y(\Gamma)$ decreases slightly
with increasing $\Gamma$. The attractive-glass branch increases with
increasing $\Gamma$.
If one choses a path through the state diagram that crosses the glass--glass
transition line, these two branches will cover the full $\Gamma$ range,
but still exhibit a finite jump at the $\Gamma_c$ corresponding to the
glass--glass transition.
In the range of attraction ranges $\delta$
close to $\delta^*$ studied here,
the yield stress is almost independent on $\delta$ along the attractive-glass
branch.
For strong attraction, it can be described by $\sigma_y\sim\Gamma^y$ with
some exponent $y$ close to $2$, and
with a prefactor close to $k_BT/d^3$. For our data, $y\approx2.2$ gives a
good fit. Similar behavior has also been predicted
by \textcite{Kobelev2005} and seen in experiment, where
different power laws have been observed and attributed to an oversimplification
in the square-well model \cite{Laurati2011}.
A scaling with attraction strength with $y=2$
is expected from the linear-response
(Maxwell plateau) shear modulus $G_\infty$ in this model:
the attractive-glass line in MCT is dominated by contributions to the
coupling vertex that scale as $A^2/\delta$ \cite{Bergenholtz2000},
where $A=\exp[\Gamma]-1$. In the MSA structure factor employed in this work,
the factor $A$ arising from the Mayer cluster function is replaced by
$\Gamma$. The weak dependence of $\sigma_y$ on $\delta$ hints at the fact that
the crossover from the linear-response regime to yielding depends linearly
on the attraction range; this will be discussed below.
This crossover also depends on the attraction strength to some extent,
rationalizing why $\sigma_y$ shows a somewhat stronger dependence on $\Gamma$
as expected from $G_\infty$.
For the larger $\delta$, where the chosen line of constant packing fraction
remains in the glass at all $\Gamma$, the repulsive- and attractive-glass
branches of the $\sigma_y$-versus-$\Gamma$ curves in Fig.~\ref{syGamma}
join smoothly, 
The case $\delta=0.09$ is close to the point where the reentrant glass
transition vanishes. Concomitantly, the yield stress in this case starts
to increase monotonically with increasing $\Gamma$.

\begin{figure}
\includegraphics[width=.9\linewidth]{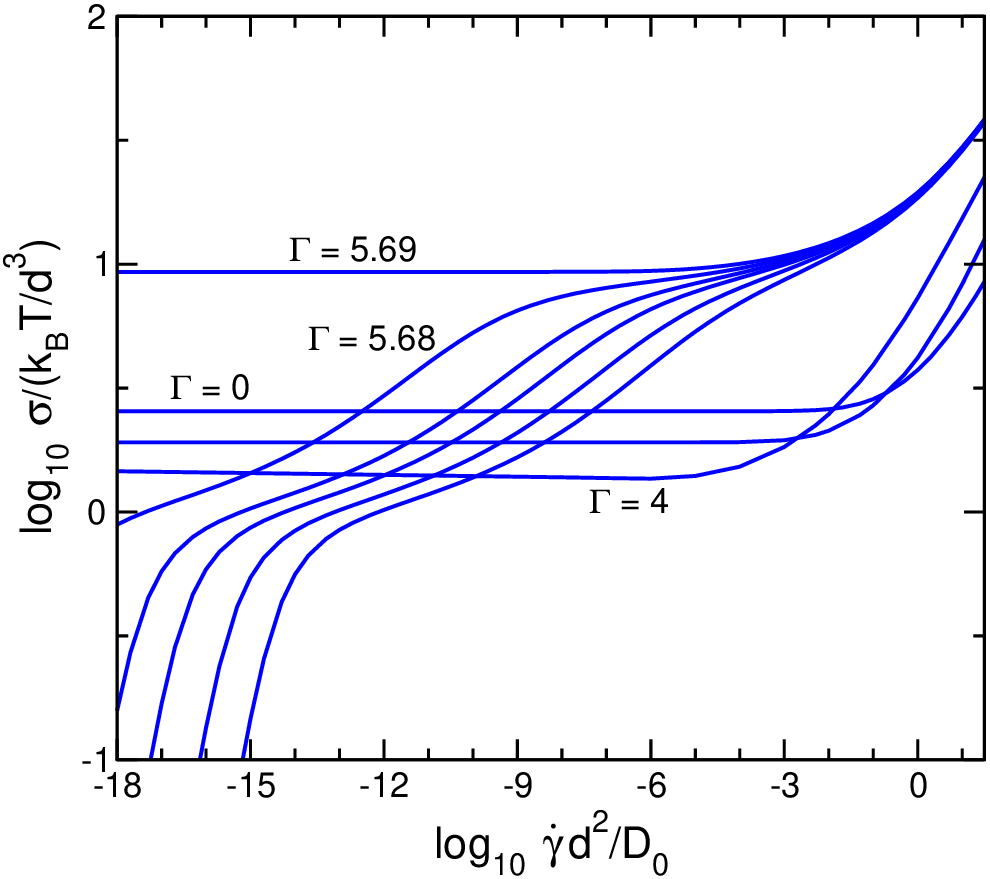}
\caption{\label{flowcurve}
  Flowcurve $\sigma(\dot\gamma)$ for the steady-state shear stress as a
  function of shear rate, for the square-well system with $\delta=0.03$,
  at fixed packing fraction $\varphi=0.535$, for increasing attraction
  strength (from bottom to top at large $\dot\gamma$: $\Gamma=0$, $2$, $4$,
  $5.61$, $5.64$,
  $5.66$, $5.67$, $5.68$, and $5.69$).
}
\end{figure}

The yield-stress plateau in the flow curves becomes the more pronounced,
the closer the system is to the respective glass transition.
Thus, for states close to a glass--glass transition, flowcurves with
an indication of two plateaus, corresponding to $\sigma_y^\text{rep}$ and
$\sigma_y^\text{attr}$, can be expected. The flowcurves are monotonically
increasing with increasing $\dot\gamma$, hence the repulsive-glass yield-stress
plateau will appear at lower $\tlname{Pe}$, and the attractive-glass one
at larger $\tlname{Pe}$. This is shown in Fig.~\ref{flowcurve} for the
representative case $\delta=0.03$ and $\varphi=0.535$, for increasing attraction
strength.
In order to demonstrate the qualitative signature of the two glasses
that emerges, a
range of shear rates has to be shown that is unrealistically large for typical
experiments or simulations.

As in Fig.~\ref{d04flowcurve}, the hard-sphere reference case (dashed line
in Fig.~\ref{flowcurve}) exhibits a yield stress of the order $k_BT/d^3$.
In Fig.~\ref{flowcurve} we do not focus on the weak-attraction regime
since this is qualitatively the same as above. For the relatively strong
attractions close to the glass--glass transition line, the repulsive-glass
yield stress is already lowered appreciably, due to the softening of the
glass connected to the reentrant fluidization. This is exhibited by the
lower-$\Gamma$ curves in Fig.~\ref{flowcurve}: there, a repulsive-glass
yield-stress plateau of $\mathcal O(k_BT/d^3)$ is visible around
$\tlname{Pe}\approx10^{-12}$, that is already well below the yield stress
calculated for the hard-sphere system at this packing fraction.
Since at $\varphi=0.535$, these curves fall into the reentrant-fluidized
region, they all display a linear Newtonian behavior at even lower
$\tlname{Pe}$.

The curve for $\Gamma=5.68$ in Fig.~\ref{flowcurve} demonstrates the
approach to the glass--glass transition. The incipient arrest driven
by attraction results in an additional, larger attractive-glass yield-stress
plateau at higher (but still small) $\tlname{Pe}\approx10^{-6}$. A flowcurve
with two plateaus results.
Within a window of shear rates that is realistically achievable in
experiment or simulation, this distinction is not clearly visible.
Rather, a broad window of sublinear rise in shear stress $\sigma$ as
a function of shear rate $\dot\gamma$ is characteristic for states
close to the glass--glass transition, as was already discussed in connection
with Fig.~\ref{d04flowcurve}.

\begin{figure}
\includegraphics[width=.9\linewidth]{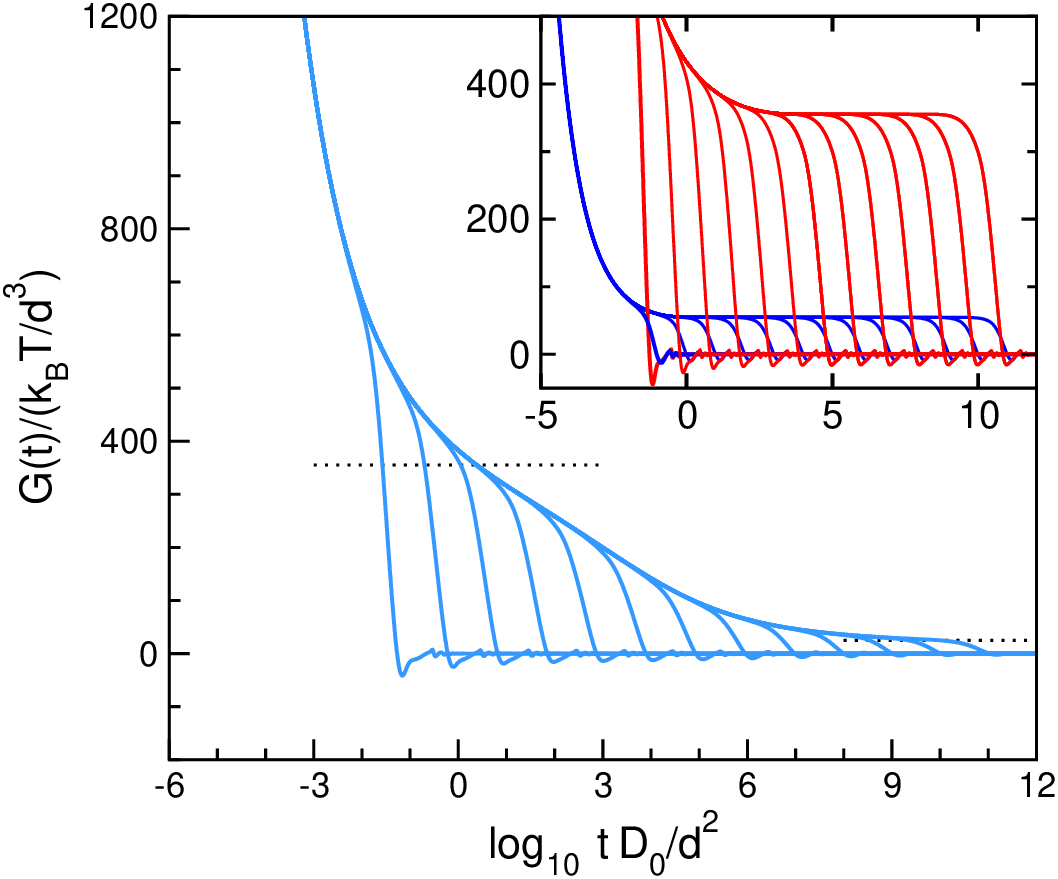}
\caption{\label{modulus}
  Dynamical shear modulus $G(t)$ as a function of time $t$ for the SWS
  with $\delta=0.03$, packing fraction $\varphi=0.535$, and attraction
  strength $\Gamma=5.64$, close to the glass--glass transition. Various
  shear rates, $\dot\gamma\tau_0=10^n$ with $n=-12,-11,\ldots,0$, are
  shown. Dotted lines indicate the plateau moduli of the nearby attractive
  and repulsive glasses.
  The inset shows the corresponding plot for $\Gamma=5.69$,
  inside the attractive glass (upper set of curves), and for the repulsive
  glass, $\Gamma=0$ (lower set of curves).
}
\end{figure}

The behavior discussed above for the flowcurves reflects the non-trivial
decay of the ITT-MCT correlation functions close to the glass--glass
transition. Since the attractive glass is characterized by a larger
nonergodicity parameter $f_q$ than the repulsive glass, the quiescent density
correlation functions are asymptotically governed by two plateaus:
at short times, they relax to the larger $f_q^\text{attr}$, where they
arrest in the attractive glass for sufficiently large $\Gamma$.
For smaller $\Gamma$, they decay from this
plateau towards the lower $f_q^\text{rep}$. For state points in the liquid,
there then is a final relaxation from this plateau to zero.
The intricacies of this three-step relaxation (as opposed to the common
two-step relaxation in the vicinity of ordinary glass transitions)
will not be discussed here in detail; we refer to the literature
\cite{Fuchs1991b,Dawson2001,Sperl2004}.

The asymptotic reasoning carries over to the dynamical shear modulus
$G(t)$, Eq.~\eqref{gt}. Essentially, one can distinguish three different
scenarios for the decay of $G(t)$, depending on the ratio of the shear rate
to the two relaxation times involved in the decay first from the
attractive-glass plateau, $\tau^\text{attr}$, and second from the
repulsive-glass plateau, $\tau^\text{rep}\gg\tau^\text{attr}$.
One can then introduce two different dressed P\'eclet numbers,
$\tlname{Pe}^\text{rep}=\dot\gamma\tau^\text{rep}$ and
$\tlname{Pe}^\text{attr}=\dot\gamma\tau^\text{attr}$.
For $\tlname{Pe}^\text{attr}\ll\tlname{Pe}^\text{rep}\ll1$, one
recovers the quiescent equilibrium curve for $G(t)$, and hence linear
response with a Newtonian shear viscosity. Increasing the shear rate,
one enters a regime $\tlname{Pe}^\text{rep}\gg1$ but
$\tlname{Pe}^\text{attr}\ll1$, where the stress is asmymptotically
given by the yield stress of the repulsive glass. Finally, in the
regime $\tlname{Pe}^\text{rep}\gg\tlname{Pe}^\text{attr}\gg1$,
the stress is given by the yield stress of the attractive glass, until
short-time relaxation effects become dominant at the highest shear
rates.
Since separating the two plateau regions in the correlation functions
requires to fine-tune the parameters to the (liquid-side of the)
crossing point of the two glass-transition lines in Fig.~\ref{gt-diagram},
the asymptotic behavior described above is unlikely to be clearly
observed.
This is seen in Fig.~\ref{modulus}, where we plot in the main panel
the dynamical shear modulus for the attraction strength $\Gamma=5.64$
also considered in Fig.~\ref{flowcurve},
and various shear rates, exemplifying the case
close to the crossing of glass-transition lines. For $\dot\gamma\to0$
these curves approach the quiescent correlation function that exhibits
indications of a three-step relaxation process involving two plateaus,
and a relaxation time $\tau^\text{attr}\sim10^3\,D_0/d^2$. Although
this relaxation time is already very large, there still is no clearly
recognizable attractive-glass plateau in $G(t)$.
This is predicted by MCT for state points in the vicinity of higher-order
transition points (such as the endpoint of the glass--glass transition
line). There, the asymptotic description of two-step relaxation with a
power law towards the plateau and a power-law decay from this plateau,
is replaced by logarithmic decay laws \cite{Goetze2002}, visible in
semi-log plots such as Fig.~\ref{modulus} as straight lines.
As shown in the inset of the figure, the dynamical shear moduli for the
repulsive glass at $\Gamma=0$, and the attractive glass at
$\Gamma=5.69$, clearly show the emergence of the related plateaus as
the shear rate is lowered.

Shear induces a decay in $G(t)$ on a time scale $\tau_{\dot\gamma}$ that
is for all states determined by the shear-flow time scale $1/\dot\gamma$.
This decay fully decorrelates the fluctuations
in ITT-MCT, so that no trace of the lower repulsive-glass plateau in $G(t)$
remains for large shear rates.

\begin{figure}
\includegraphics[width=.9\linewidth]{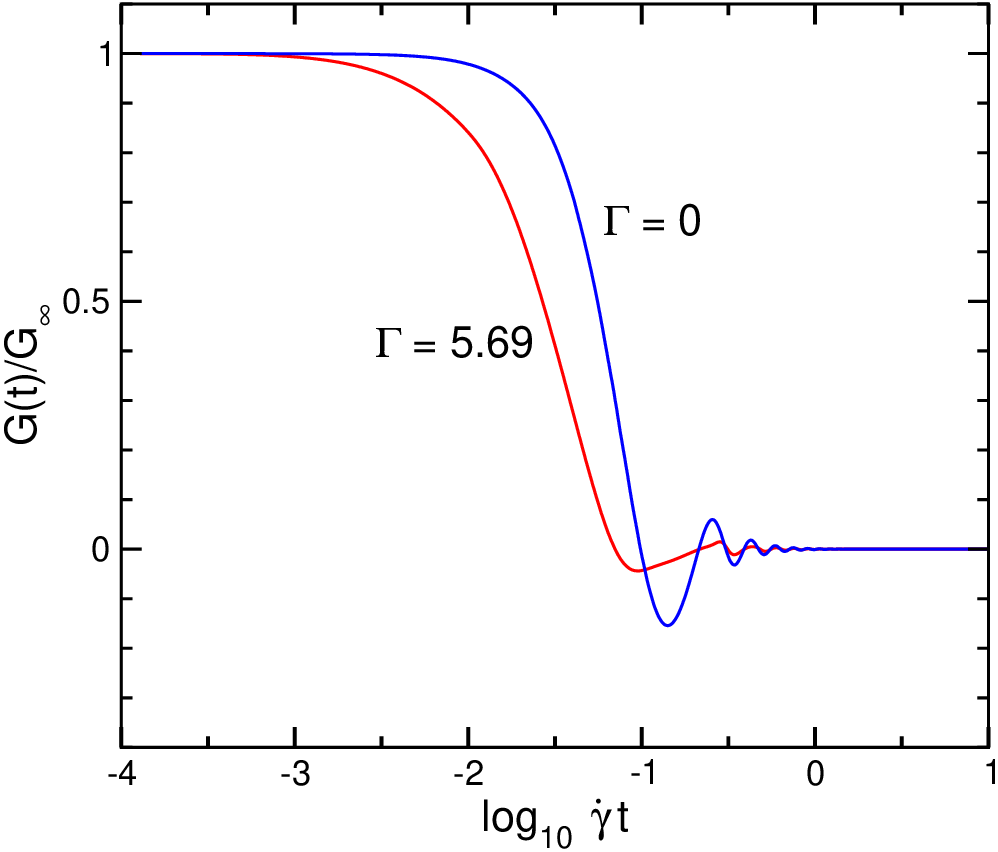}
\caption{\label{modscaled}
  Yield-stress master curves for the dynamical shear modulus
  of the SWS at $\delta=0.03$ and $\varphi=0.535$,
  for $\Gamma=0$ (repulsive glass), and
  $\Gamma=5.69$ (attractive glass), normalized by the
  plateau modulus $G_\infty$, as a function of scaled time $\dot\gamma t$.
}
\end{figure}

The $G(t)$ curves for large P\'eclet numbers show negative dips at late times
in the relaxation. This has been
discussed extensively in repulsive glasses \cite{Fuchs2003,Zausch2008,Amann2013}.
Since the time integral over $G(t)$ determines the
shear stress, this corresdponds to a nonmonotonic approach to the
steady-state value $\sigma$ and reflects the existence of a stress overshoot
in the stress--strain curves measured after startup of steady flow.
Since $G(t)$ is the autocorrelation function of microscopic stresses,
this dip reflects an ``elastic recoil'' of breaking cages
\cite{Koumakis2012,Siebenbuerger2012}.

The scaling of $\tau_{\dot\gamma}$ with $1/\dot\gamma$ as the only relaxation
time that remains in the flow-melted ideal glass, implies that a master
curve for the relaxation is obtained by scaling $G(t)$ by the plateau
value $G_\infty$ and time by $\dot\gamma$.
Figure~\ref{modscaled} shows master curves obtained from the
$\dot\gamma=10^{-12}$ curves for the repulsive glass
and the attractive glass at $\varphi=0.535$ and $\delta=0.03$.
While the asymptotic laws of ITT-MCT describe generic behavior of the
correlation functions around their plateau value, the details of the
decay from this plateau to zero depend on the details of the interaction
potential and state points. As exhibited in Fig.~\ref{modscaled}, the
repulsive glass features a steeper decay to zero than the attractive-glass
master curve for the value of $\Gamma$ shown. Additionally, the
coefficient relating $\tau_{\dot\gamma}$ to $1/\dot\gamma$ decreases with
increasing attraction strength. In the repulsive glass, we obtain
$\tau_{\dot\gamma}\approx\gamma_c/\dot\gamma$, in agreement with
previous results \cite{Fuchs2003,Zausch2008}.
Recall that the stress--strain
curves are given by $\sigma_{xy}(\gamma)=\dot\gamma\int_0^tG(t)\,dt$ for
$\gamma=\dot\gamma t$.
This implies that
the overshoot in the stress--strain curves of the repulsive glass
occurs at a strain of roughly $\gamma_c$ and corroborates our choice
of $\gamma_c=0.1$ to reflect the Lindemann criterion.
In the attractive glass, the decay in $G(t)$ sets in earlier.
This implies that the position of the overshoot in the stress--strain
curves moves to lower strains with increasing $\Gamma$, as will be discussed
in detail below.

\begin{figure}
\includegraphics[width=.9\linewidth]{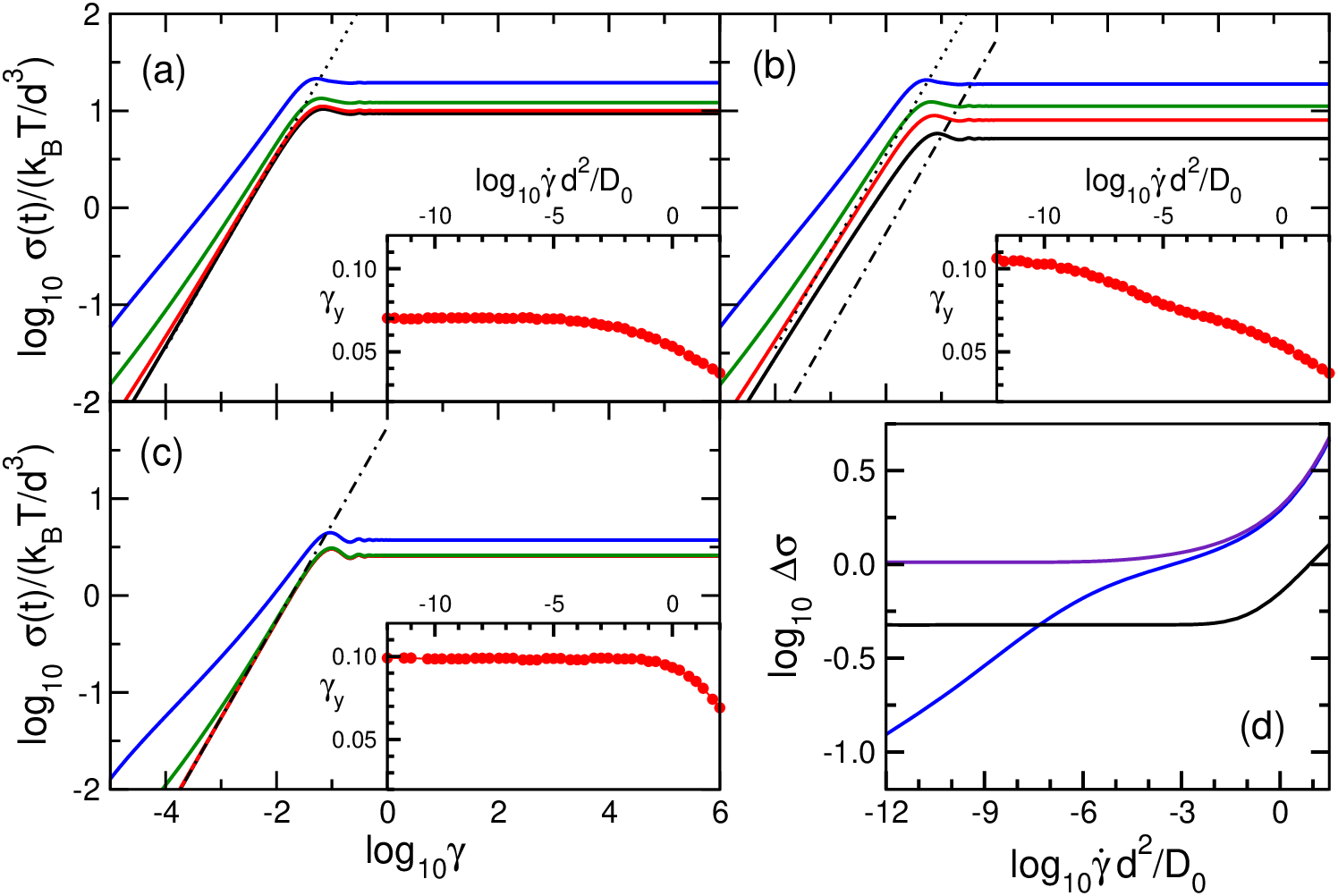}
\caption{\label{stressstrain}
  Stress--strain curves after startup of steady shear, $\sigma(\gamma)$
  for the cases shown in Fig.~\ref{modulus} and various shear rates
  $\dot\gamma\tau_0=10^n$ with $n=-6$, $-4$, $-2$, and $0$ (bottom to top):
  (a) for the attractive-glass case, $\Gamma=5.69$,
  (b) close to the glass--glass transition, $\Gamma=5.64$,
  and
  (c) for the repulsive hard-sphere glass, $\Gamma=0$.
  Dotted (dash-dotted) lines mark the linear-elastic plateau moduli
  $G_\infty$ for the attractive (repulsive) glass cases.
  Insets show the dependence of the overshoot position
  $\gamma_y$ as a function of shear rate.
  (d) Overshoot strength
  $\Delta\sigma=(\sigma_\text{max}-\sigma)/\sigma_\text{max}$ as
  a function of shear rate, for various $\Gamma$.
}
\end{figure}

We investigate the stress overshoots in more detail in
Fig.~\ref{stressstrain}. There, stress--strain curves $\sigma(\gamma)$
after startup of steady shear with various shear rates $\dot\gamma$ are
shown for three representative cases: the repulsive glass, the attractive
glass, and a state point close to the glass--glass transition.
All curves show similar stress overshoots, showing that this feature
is a robust signature of cage breaking, be it induced by the breaking
of nearest-neighbor topology, or the breaking of individual bonds.
The stress--strain curves for low $\tlname{Pe}$ all show a linear elastic
regime at small strains that is governed by the plateau value of the
shear modulus, $G_\infty$. The corresponding values obtained from the
cases $\Gamma=0$ and $\Gamma=5.69$ are indicated in the figures as
dash-dotted and dotted lines (cf.\ Fig.~\ref{modulus}). At large
$\tlname{Pe}$, shear-induced decay starts to set in already during the
decay of $G(t)$ towards $G_\infty$, and the elastic regime in $\sigma(\gamma)$
starts to correspond to the much higher short-time modulus $G_0$. (The
latter is formally infinite for true hard spheres.) This regime will not
be discussed here.
For the intermediate case close to the glass--glass transition, still
lower shear rates would be required to probe the repulsive-glass
linear-response $G_\infty$.

It should be noted that none of our stress--strain curves indicate the
rich behavior at very large strain, $\gamma\approx1$, that was seen
in experiment \cite{Pham2006,Pham2008,Laurati2009,Koumakis2011}. Instead of a double-yielding
scenario, ITT-MCT predicts both the repulsive and the attractive glass to
yield at strains small compared to unity. In principle, the three-step decay
of $G(t)$ induces several shoulders in the low-$\gamma$ part of the
stress--strain curve, but this concerns strains that are far below the
experimental resolution, only probed at extremely small startup shear
rates.

The mechanisms captured by ITT-MCT are more clearly borne out in the
corresponding yield strain, i.e., the position of the maximum, $\sigma(\gamma_y)
=\sigma_\text{max}$. The dependence of $\gamma_y$ on shear rate is
shown in the insets of the first three panels in Fig.~\ref{stressstrain}.
The strong dependence that sets in at the highest shear rates shown are
indicative of the cross-over to large bare P\'eclet numbers, and cannot
be discussed correctly within the present ITT-MCT.
Considering low shear rates for the repulsive glass, the stress overshoot
occurs at a roughly rate-independent strain of $\gamma_y^\text{rep}\approx0.1$,
recovering the Lindemann criterion as discussed above. Since this value
is essentially set by the model parameter $\gamma_c$ in the isotropic
approximation to ITT-MCT, it is not predictive.

In the attractive glass, we find $\gamma_y^\text{rep}\approx0.07$ for the
state point chosen here, again weakly dependent on shear rate for low
enough $\tlname{Pe}$. This reproduces an effect discussed previously in
experiment \cite{Koumakis2011}: the (first) yield strain of the attractive
glass is lower than that of the repulsive glass. In the vicinity of the
glass--glass transition, as explained above, the shear rate controls whether
yielding is dominated by the attractive- or by the repulsive-glass shear
modulus. Indeed, for the $\Gamma=5.64$ curves in Fig.~\ref{stressstrain},
we observe a more pronounced dependence of $\gamma_y$ on $\dot\gamma$. Its
value decreases from $\gamma_y\approx\gamma_y^\text{rep}$ at the lowest
shear rates to $\gamma_y\approx\gamma_y^\text{attr}$ at higher shear rates.
This agrees with the discussion of $G(t)$ above.

The relative strength of the overshoot as compared to the steady-state
value, $\Delta\sigma=(\sigma_\text{max}-\sigma)/\sigma_\text{max}$, is
shown in the lower right panel of Fig.~\ref{stressstrain}. For the
repulsive- and the attractive-glass limits, $\Delta\sigma$ is essentially
independent on $\dot\gamma$ in the range of shear rates that are of interest
here. This reflects the fact that within ITT-MCT, $G(t)$ approaches a
master curve that depends only on $\dot\gamma t$ for states in the glass
and in the limit $\dot\gamma\to0$, as discussed in connection with
Fig.~\ref{modscaled}.
For the ultimately
liquid state $\Gamma=5.64$ shown in Fig.\ref{stressstrain}, the
overshoot has to vanish as $\dot\gamma\to0$, since in the linear-response
regime $G(t)$ has to decay as a monotonically decreasing positive function
\cite{Siebenbuerger2012}. In effect, the $\Delta\sigma$-versus-$\dot\gamma$
curves qualitatively resemble the corresponding flow curves.

\begin{figure}
\includegraphics[width=.9\linewidth]{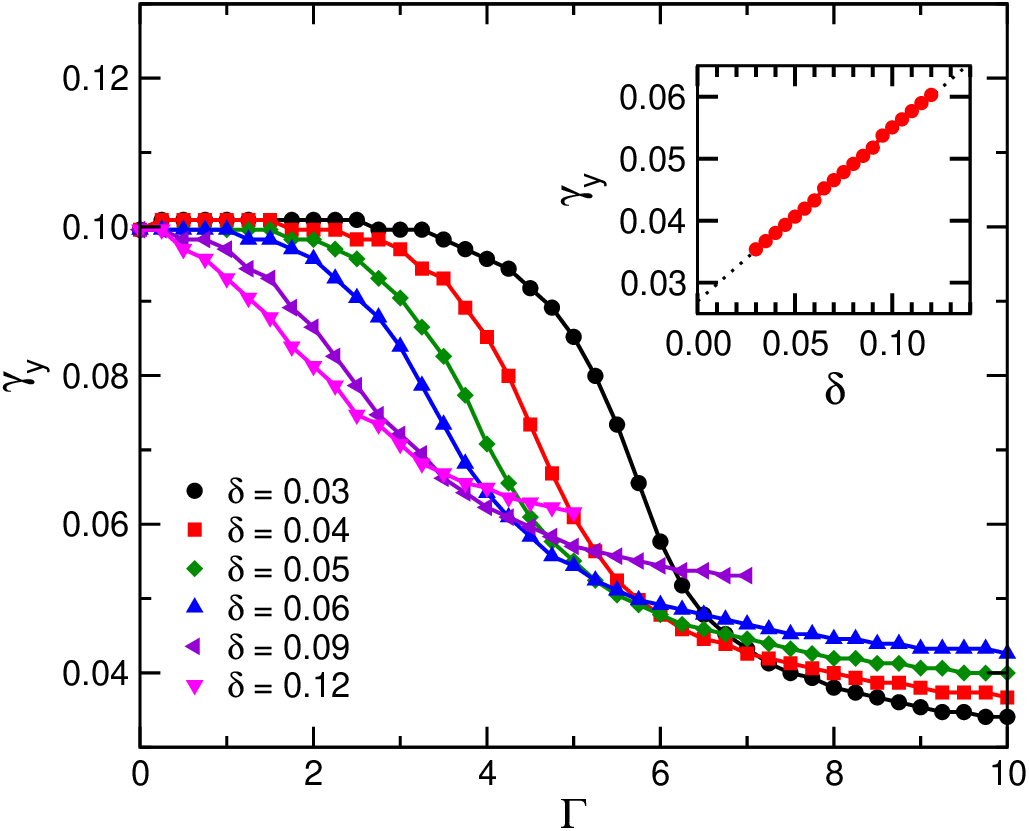}
\caption{\label{gammayGamma}
  Dependence of the yield strain $\gamma_y$ on the attraction strength
  $\Gamma$ for various attraction ranges $\delta$ as labeled, for
  strong shear, $\tlname{Pe}=0.1$.
  The inset shows the dependence of the large-$\Gamma$ asymptote
  on $\delta$. A dotted line indicates a linear relation.
}
\end{figure}

Let us discuss in more detail the dependence of the (first) yield strain
$\gamma_y$ on the parameters of the square-well attraction.
Figure~\ref{gammayGamma} shows $\gamma_y(\Gamma)$ for various $\delta$.
The trend indicated already above becomes clear here: for each attraction
range $\delta$, there is a crossover from a repulsive-glass-like yielding
at $\gamma_y^\text{rep}\approx0.1$ at low attraction strength $\Gamma$, to a
strong-attraction asymptote where $\gamma_y^\text{attr}<0.1$. The value of
$\Gamma$ where this crossover occurs, increases with decreasing $\delta$.
For low $\delta$, this can be rationalized by a concommitant trend in
the glass--glass transition line which also moves to stronger attraction
with decreasing square-well width.

The large-$\Gamma$ asymptote obeys $\gamma_y^\text{attr}\propto\delta$ with
some offest,
as shown in the inset of Fig.~\ref{gammayGamma}.
This is in line with the reasoning that the yield
strain in the attractive glass is eventually set by the length of a bond.
It should however be noted that the constant of proportionality is less than
unity. Due to the offset, $\gamma_y^\text{attr}(\delta\!=\!0)>0$ holds, but
recall that the large-$\Gamma$ asymptote in this limit would only be obtained
for $\Gamma\to\infty$, i.e., Baxter's adhesive-sphere limit \cite{Fabbian1999,Dawson2001}. For small $\delta$, we thus obtain $\gamma_y^\text{attr}\gtrsim\delta$
(cf.\ the case $\delta=0.03$ in our calculations). But already for slightly larger attraction ranges, $\gamma_y^\text{attr}\lesssim\delta$. In particular, up to
$\delta=0.12$ we still find $\gamma_y^\text{attr}<\gamma_y^\text{rep}$ and
$\gamma_y^\text{attr}\sim\delta$, while
the argument based on the Lindemann length scale would suggest that at such
large $\delta$, the yielding of cages should come first.
One should however recall that the Lindemann length is only a rough estimate,
and that this cage-induced length scale will also decrease upon entering
deeper into the glass.
Let us also note in this context,
that at large $\Gamma$, one approaches the spinodal
of the SWS model, which may bring in additional physical mechanisms. We have
restricted all data points to be sufficiently far from the spinodal to not
be qualitatively influenced by it.

The two trends described above lead to a crossing of the
$\gamma_y$-versus-$\Gamma$ for different $\delta$. This in turn implies
that for weak attraction strength, there is a regime where $\gamma_y$
decreases with increasing attraction range, or even shows nonmonotonic behavior
with $\delta$. In this regime, $\gamma_y$
is not indicative of the breaking of bonds, but rather reflects the
easier fluidization of the weakly attractive glass, as compared to the
hard-sphere reference.

\begin{figure}
\includegraphics[width=.9\linewidth]{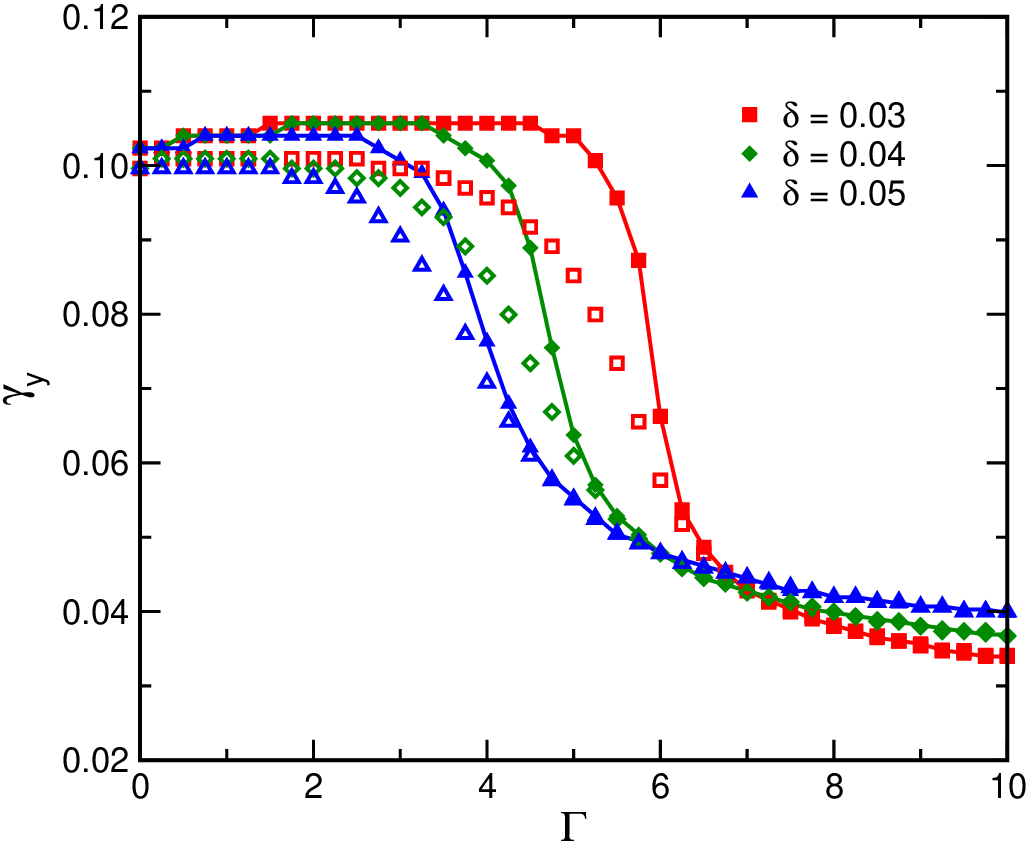}
\caption{\label{gammayGammalow}
  As in Fig.~\ref{gammayGamma}, but for a lower shear rate
  $\tlname{Pe}=10^{-3}$ (filled symbols). Open symbols repeat the data
  of Fig.~\ref{gammayGamma} ($\tlname{Pe}=0.1$) as a reference.
}
\end{figure}

The position of $\gamma_y$ depends weakly on the shear rate, and it does
most noticably so in the intermediate-$\Gamma$ regime close to the
glass--glass transition (cf.\ insets in Fig.~\ref{stressstrain}).
Hence, the $\gamma_y$-versus-$\Gamma$ curves may display additional
behavior in the crossover regime for different shear rates. This is
exemplified in Fig.~\ref{gammayGammalow}, where results corresponding to
those shown in Fig.~\ref{gammayGamma} are repeated for lower $\tlname{Pe}$.
For large $\Gamma$, the overshoot position does not depend strongly on
$\tlname{Pe}$. Hence, the curves for same $\delta$ agree in this limit.
But for the $\delta\lesssim\delta^*$ curves displayed in the
figure, a weak non-monotonic behavior
in $\gamma_y(\Gamma)$ is observed at intermediate $\Gamma$:
Up to $\Gamma\approx3$, the yield
strain for, say, $\delta=0.03$ and
$\tlname{Pe}=10^{-3}$ first increases slightly with increasing
$\Gamma$, before it decreases to the expected large-$\Gamma$ asymptote.
An increase of $\gamma_y$ with increasing $\Gamma$ has also been
observed in full anisotropic ITT-MCT calculations
\cite{AmannJOR}. For the shear rate shown in Fig.~\ref{gammayGammalow}, we
only find nonmonotonic behavior in $\gamma_y(\Gamma)$ for small enough $\delta$,
but not for $\delta=0.09$ or $\delta=0.12$ (not shown).

The mechanism of shear-advected loss of contributions to the MCT coupling
coefficients allows to qualitatively explain the initial
increase in $\gamma_y$ with increasing $\Gamma$: adding weak attractions to
the hard-sphere system, the first effect observed in the static structure
factor $S_q$ (and thus similarly in the DCF $c_q$) is a lowering of the
peak amplitudes, combined with an increase in peak widths \cite{Dawson2001}.
The lowering of amplitude is responsible for the attraction-induced weakening
of the glass. Yielding within ITT-MCT occurs at the point where the accumulated
strain is sufficient to decorrelate the advected and non-advected contributions
to $V_{\vec q,\vec k}^{(\dot\gamma)}$, cf.\ Eq.~\eqref{vqk}. For those
regions where the broadening of the peaks in $c_q$ is still dominant over
its overall loss of amplitude, the decorrelation will require larger strains
than in the hard-sphere reference state. Hence, $\gamma_y(\Gamma)$ initially
increases, as shown in Fig.~\ref{gammayGammalow}. For stronger attractions,
the growing large-$q$ tail of $c_q$ becomes more dominant. The attraction
range $\delta$ in this case governs the large-$q$ periodicity of the
DCF, so that decorrelation in Eq.~\eqref{vqk} is achieved by strains that
themselves scale with $\delta$.

The finding that $\gamma_y\sim\delta$ independent on
$\Gamma$ in the strong-attraction limit, provides a qualitative explanation
for the independence of the yield stress on $\delta$ that was discussed
in connection with Fig.~\ref{syGamma}. Noting that the shear-induced
decay in $G(t)$ is exponential or faster, the integral determining the
yield stress is dominated by the area under the platau of height $G_\infty$,
cut off at the time where $\dot\gamma t\approx\gamma_y$. As a result,
$\sigma_y\sim G_\infty\gamma_y$. Recalling the above-mentioned scaling of
$G_\infty$, one gets $\sigma_y\sim(\Gamma^2/\delta)\cdot\delta$, which
is independent on $\delta$.

\begin{figure}
\includegraphics[width=.9\linewidth]{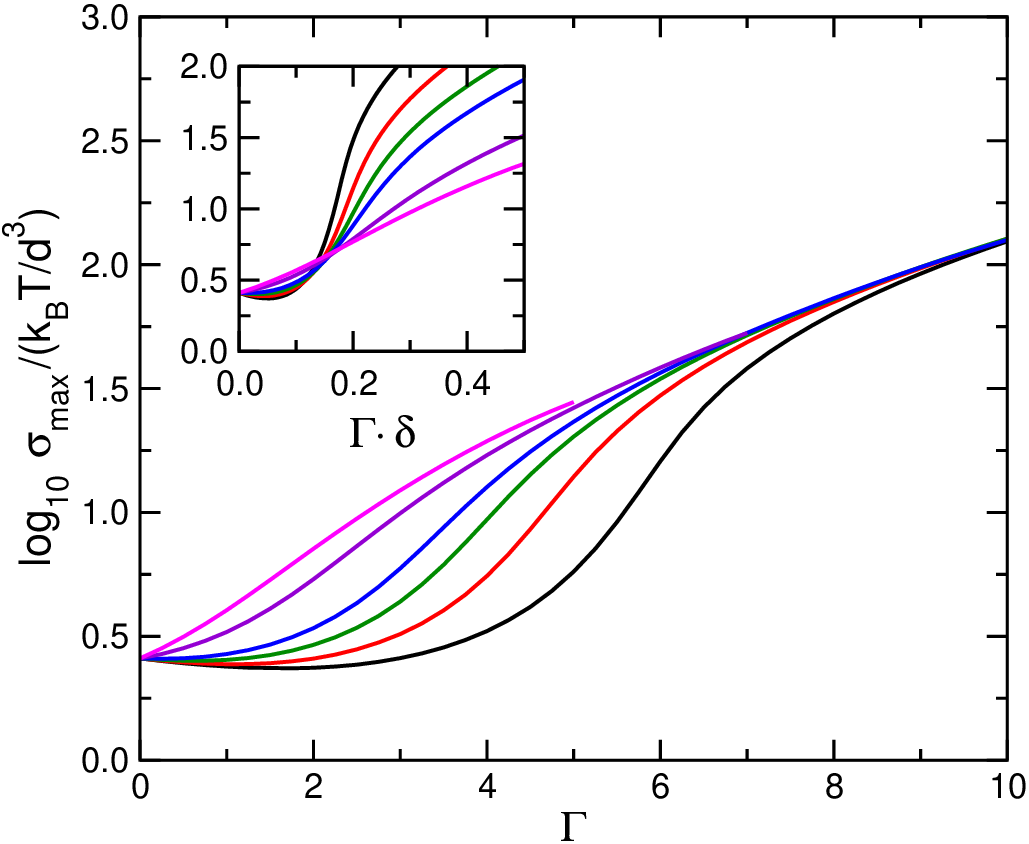}
\caption{\label{smaxGamma}
  Dependence of the maximum stress $\sigma_\text{max}$ at yielding on
  the attraction strength $\Gamma$ in the square-well system, for
  packing fraction $\varphi=0.525$ and $\dot\gamma=0.1$. Various
  attraction ranges $\delta=0.03$, $0.04$, $0.05$, $0.06$, $0.09$, and $0.12$,
  are shown (increasing from bottom to top).
  The inset shows the same data versus the attraction area $\Gamma\cdot\delta$.
}
\end{figure}

The dependence of the maximum stress at the yield strain, $\sigma_\text{max}
=\sigma(\gamma_y)$, on the attraction strength $\Gamma$
is documented in Fig.~\ref{smaxGamma}. At the largest
$\Gamma$ shown, $\sigma_\text{max}$ depends only weakly on the range $\delta$,
and increases strongly with $\Gamma$. This is in qualitative agreement
with the behavior predicted for the dynamic yield stress
discussed above.
For $\sigma_\text{max}$,
the increase with $\Gamma$ is found to be slightly stronger than
$\Gamma^2$, and can be fitted with an effective power law, $\Gamma^\alpha$
with $\alpha\approx2.2$, in the range shown in Fig.~\ref{smaxGamma}.
This is similar to the behavior found for the steady-state dynamic yield
stress, cf.\ Fig.~\ref{syGamma}.
As already found for $\gamma_y$, the crossover into this strong-attraction
regime scales with the position of the glass--glass transition and hence
shifts to larger $\Gamma$ with decreasing $\delta$. This is demonstrated
in the inset of Fig.~\ref{smaxGamma}, where $\sigma_\text{max}$ is shown
versus the square-well attraction area $\Gamma\cdot\delta$.

At intermediate $\Gamma$, a minimum in the $\sigma_\text{max}$-versus-$\Gamma$
curve is seen. This is again a signature of the reentrant glass melting.
The initial decrease with increasing $\Gamma$ reflects the weakening
of the glass due to weak attraction: it agrees with the decrease
of the yield stress shown in Fig.~\ref{syGamma} and the observation
that the relative stress-overshoot height is insensitive to the
attraction parameters. Note that in Fig.~\ref{smaxGamma}, a finite
P\'eclet number, $\tlname{Pe}=0.1$, was chosen. Hence, an overshoot is
still observed in the attraction-melted fluid at intermediate $\Gamma$,
where in Fig.~\ref{syGamma}
no true yield stress could be defined.

\begin{figure}
\includegraphics[width=.9\linewidth]{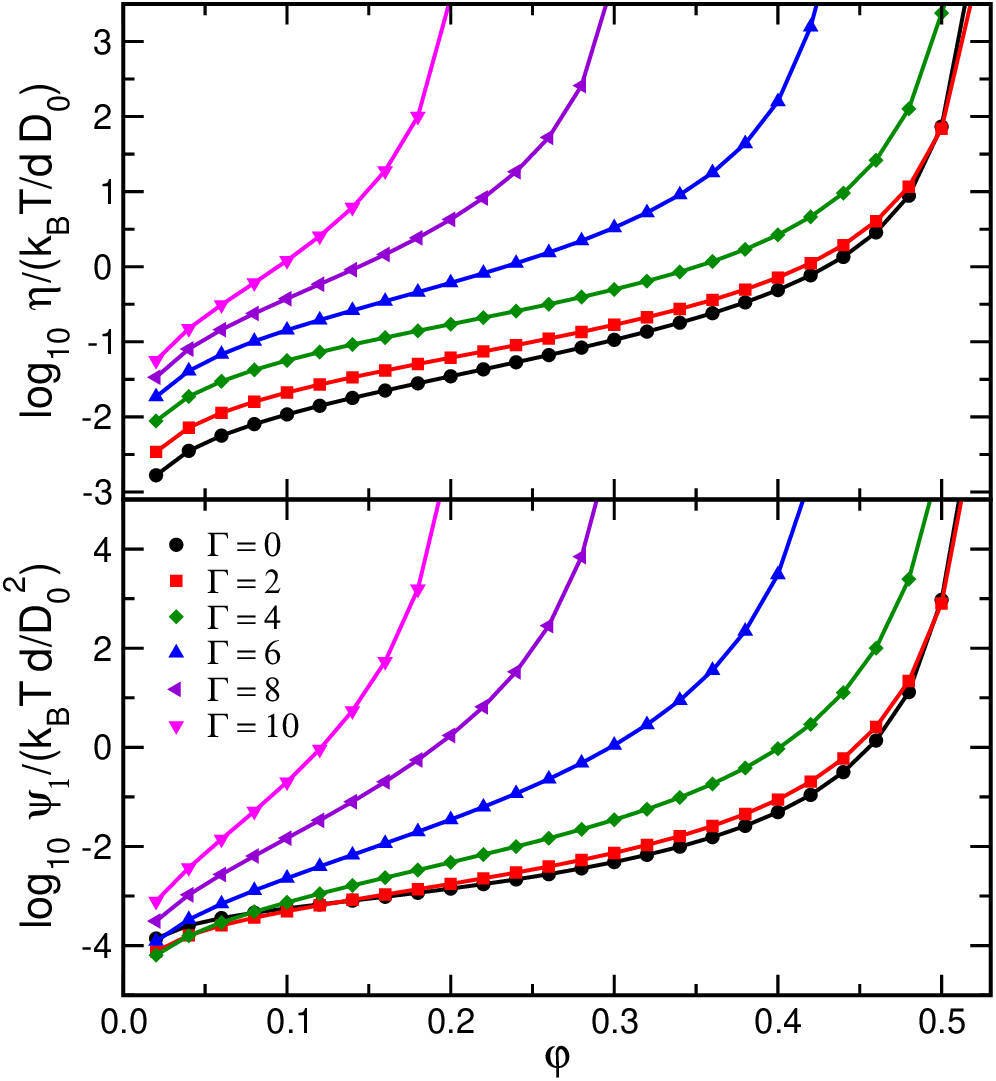}
\caption{\label{etapsi}
  Low-shear viscosity $\eta=\sigma/\dot\gamma$ (upper panel) and first
  normal-stress coefficient $\Psi_1=(\sigma_{xx}-\sigma_{yy})/\dot\gamma^2$
  (lower panel) as functions of packing fraction $\varphi$, for various
  attraction strenghts of the SWS model with $\delta=0.05$, and
  for shear rate $\dot\gamma\tau_0=10^{-4}$.
}
\end{figure}

We now turn to a brief discussion of the relation between shear viscosity
and the first normal-stress coefficient.
\textcite{Farage2013} have investigated this
relationship using the full, anisotropic ITT-MCT, in an expansion of small
shear rates. Following this work, we present in Fig.~\ref{etapsi} the
results for the shear viscosity $\eta$ and the first normal-stress coefficient
$\Psi_1$ as functions of packing fraction $\varphi$
from the isotropic model for a constant-$\Gamma$
path through configuration space.
Qualitatively, our results agree very
well with those of \textcite{Farage2013}, cf.\ their Fig.~2, although one
has to keep in mind that these authors were discussing a hard-core Yukawa
potential as a model for a short-ranged attractive colloid, instead of the SWS.
This difference notwithstanding, we confirm that the isotropic ITT-MCT
model produces very similar results to the anisotropic theory in this limit.
As the packing fraction is increased, both the viscosity and the normal-stress
coefficient increase strongly since the glass transition is approached.
This increase sets in at lower packing fractions for stronger attractions,
as expected from the glass-transition diagram, Fig.~\ref{gt-diagram}.
$\Psi_1$ is about one
order of magnitude smaller than $\eta$ at low densities, but diverges more
quickly towards the transition. This is in quantitative agreement with
the low-$\dot\gamma$ limit of the full
theory \cite{Farage2013}. There, also the second
normal-stress coefficient $\Psi_2$ was investigated. It remains approximately
one order of magnitude below $\Psi_1$ throughout. Our isotropic approximation,
which entails setting $\Psi_2=0$, hence seems justified in this respect.

\begin{figure}
\includegraphics[width=.9\linewidth]{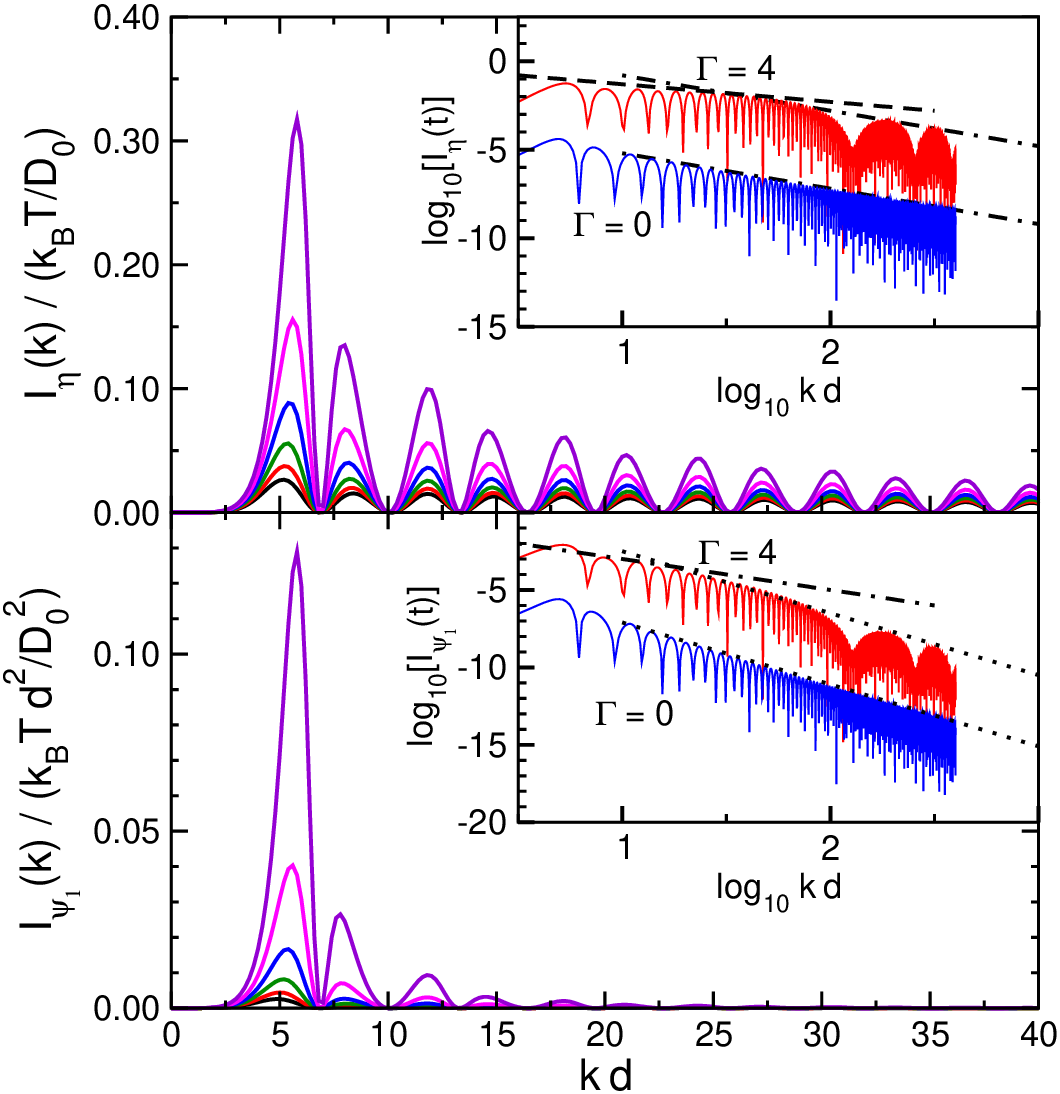}
\caption{\label{ietaipsi}
  Viscosity- and normal-stress-coefficient integrands $I_\eta(k)$ (upper panel)
  and $I_{\Psi_1}(k)$ (lower panel), as a function of wave number $k$
  in the isotropic ITT-MCT model. Different curves correspond to increasing
  packing fraction (increasing from bottom to top: $\varphi=0.16$, $0.20$,
  $0.24$, $0.28$, $0.32$, and $0.36$), for $\Gamma=4$,
  $\delta=0.05$, and $\dot\gamma\tau_0=10^{-4}$.
  The insets show the $\varphi=0.24$ results in a double-logarithmic
  representation. Dashed lines indicate $1/k$ laws, dash-dotted lines
  $1/k^2$, and dotted lines $1/k^4$ laws for comparison.
  Here, results for the purely repulsive case, $\Gamma=0$, have also been
  added, shifted down by a factor $1000$ for clarity.
}
\end{figure}

To investigate the relevant wave-number range contributing to the ITT-MCT
integrals for the shear stress and the first normal-stress coefficient,
we discuss the corresponding integrand functions $I_\eta(k)$ and
$I_{\Psi_1}(k)$. For this discussion, we consider low volume fractions,
to make contact with the work of \textcite{Farage2013}.
Our results from the isotropic ITT-MCT are shown in Fig.~\ref{ietaipsi}.
First one recognizes strong oscillations in the integrands, anti-phased
to the oscillations of the static structure factor: the latter, e.g., has
a peak at $k=k_\text{max}\approx2\pi/d$, while both $I_\eta$ and $I_{\Psi_1}$
display a minimum there. This is clear since the mode-coupling vertices entering
the Green-Kubo relation, Eq.~\eqref{gt}, contain derivatives of $S_k$.

As the density is increased, the integrand increases at all $k$, but more
strongly at $k$ below and above the position of the first diffraction peak,
$k_\text{max}$. This reflects the increase in nearest-neighbor caging.
At the same time, $I_{\Psi_1}$ decays more rapidly at large $k$ than
$I_\eta$. This is particularly clear in a double-logarithmic representation
of these functions, as shown in the insets of Fig.~\ref{ietaipsi}.
There one recognizes power-law asymptotes followed by the maxima.
In $I_\eta$, there is an intermediate regime where $|I_\eta(k)|\sim1/k$
for $1/d\ll k\ll1/\delta$ (dashed lines in the inset). For $k\gg1/\delta$,
$|I_\eta(k)|\sim1/k^2$ is observed, modulated by oscillations whose
periodicity is set by the smaller length scale $\delta$. This crossover
from slow $1/k$ decay to a faster $1/k^2$ decay reflects the same
crossover occuring in the DCF, and discussed in detail for the
physics of the quiescent SWS-MCT \cite{Dawson2001}. With decreasing attraction
range, the slow $1/k$ decay extends further out in reciprocal space, indicating
that for very short-ranged attractive glasses, features in the mode-coupling
vertex at large $k$ remain important, while close to the repulsive glass,
the $k$ range around $k_\text{max}$ is the dominant one. This has also been
found in computer simulation \cite{Puertas2005}.
This is of course simply the observation
that the Fourier transform reflects the discontinuities of the SWS potential.

Interestingly, the normal stresses are less dominated by large $k$. This
is evident from Fig.~\ref{ietaipsi}, where the corresponding envelopes
for $|I_{\Psi_1}|$ are found to closely match $\sim1/k^2$ and $\sim1/k^4$,
respectively.
The same effect is present in the work of \textcite{Farage2013}, although
there it was not commented upon.
A simple argument allows to qualitatively understand the faster decay for
the normal-stress integrand, starting from the wave-vector
dependent integrands defining $I_\eta(k)$, Eq.~\eqref{ieta}, and
$I_{\Psi_1}(k)$, Eq.~\eqref{ipsi}. These time integrals contain a long-time
cutoff due to the shear-induced decay of the density correlation function,
on a time scale $\sim1/\dot\gamma$, modulated by $k$-dependent prefactors
arising from the advected structure functions.
Suppose this can be captured by a $k$-dependent effective cutoff time scale
$\tau(k)/\dot\gamma$.
Then, up to trivial prefactors,
$\dot\gamma I_\eta(k)\approx\dot\gamma
\int_0^\infty dt\,\exp[-\dot\gamma t/\tau(k)]=\tau(k)$, while
$\dot\gamma^2I_{\Psi_1}(k)\approx\dot\gamma^2\int_0^\infty dt\,t\,
\exp[-\dot\gamma t/\tau(k)]=\tau(k)^2$. Hence, the envelopes determining
the large-$k$ decay of $I_\eta(k)$, $1/k$ and $1/k^2$, translate approximately
into $1/k^2$ and $1/k^4$ for $I_{\Psi_1}(k)$.
As a result, the dependence of the normal-stress difference $N_1$ on
the attraction strength is weaker than that of the shear stress $\sigma$.
This is exemplified by Fig.~\ref{syGamma}. There, $N_1$ displays a
weaker increase with $\Gamma$ in the attractive-glass regime; it is
well described by $N_1\propto\Gamma$, as compared to the stronger
$\sigma_y\propto\Gamma^2$.

We also note that in \textcite{Farage2013}, a very prominent low-$k$ structure
was seen in $I_\eta(k)$ (but less so in $I_{\Psi_1}(k)$). Such a strong
peak at $k\ll2\pi/d$ is not present in our results. This reflects the
different treatment of wave-vector advection between the isotropic and
the anisotropic ITT-MCT, resulting also in a different definition
of the integrand functions that are discussed. It may also be due to
the fact that in \textcite{Farage2013}
states closer to the liquid--gas spinodal were considered.

\section{Discussion}\label{discussion}

The ITT-MCT model predicts the yield stress to increase by roughly a factor
of ten upon crossing over from the repulsive glass to the attractive glass,
for packing fractions close to the repulsive-glass transition, and for
size ratios close to the one where glass--glass transitions first emerge.
The precise value will depend on the interaction details, size polydispersity,
and the packing fraction, but qualitatively, experimental
data by \textcite{Pham2008} agree with this result, as shown in
Fig.~\ref{d04flowcurve}. Regarding these data, the authors noted that
in the attractive glass, no clear yield-stress plateau was observed even for
the lowest $\dot\gamma$ reached in the experiment. As shown in
Fig.~\ref{d04flowcurve}, this may be because the attractive glass is not
quite reached, and an intermediate sublinear crossover from the
higher attractive-glass stress probed at high $\tlname{Pe}$, to the
repulsive-glass stress probed at low $\tlname{Pe}$ covers a broad window
of shear rates. However, one also has to keep in mind, that the attractive
glass as predicted by MCT is a theoretical idealization.
Activated processes, more dominant than in the repulsive-glass regime
\cite{Sciortino2003}, as well as aging effects that are excluded from our
theoretical description may be responsible for the disappearance of a
clear attractive-glass yield-stress plateau.

In the vicinity of glass--glass transitions, the nonlinear rheology
of short-range attractive systems should in principle reflect the
intricate multi-step relaxation patterns of the quiescent dynamics.
Close to the
higher-order singularities of MCT, one indeed expects flow curves that
have logarithmic corrections, due to the logarithmic asymptotic laws
for the MCT correlation functions close to these points.
However, the distinction between attractive and repulsive glass is only
an asymptotic one at intermediate attraction strength; mapping the approach to
this asymptote will require to cover unfeasibly large windows in shear
rate, and to fine-tune model parameters. In typical experiments, the
physics of the glass--glass transition will hence rather manifest itself
in preasymptotic features, such as the effective non-trivial shear-thinning
power laws discussed in Fig.~\ref{d04flowcurve}.

The ITT-MCT yield stress of the repulsive glass is given by the entropic
energy-density scale, $\sigma_y^\text{rep}\sim k_BT/d^3$.
In the attractive glass, the attraction strength $\Gamma$ takes over
the role of the energy scale, leading to
$\sigma_y^\text{attr}\sim k_BT\Gamma^2/d^3\sim U_0(U_0/k_BT)$.
This dependence on the attraction strength reflects that of the linear-response
shear moduli. It provides an interesting perspective on recent discussions
of low-temperature glass rheology and the crossover from entropic glass
transitions to ``athermal'' ones driven by interparticle interaction energies
\cite{Ikeda2013}. Our calculations demonstrate how non-entropic yield
stresses may emerge from ITT-MCT.

Since $\sigma_y^\text{attr}$ is insensitive to the attraction range,
this indicates that the ITT-MCT yielding criterion is one of
strain-induced energy: the energy at which the glass yields is related
to the energy of the interparticle attraction, i.e., a certain number of
bonds that are broken by the deformation, and less crucially the amount
of strain needed to break such bonds.

Our isotropic-ITT-MCT calculations do not capture
the two-step yielding that was emphasized in
experiments of \textcite{Pham2006,Pham2008}, and later confirmed
\cite{Laurati2009,Koumakis2011,Koumakis2012b}. The delicate balance between attraction-
and repulsion-dominated contributions to the dynamical shear modulus concerns,
in the case where both of them are seen, very small strains that are not
resolved in experiment. Within ITT-MCT, strains of $\gamma\approx1$ are
already well into the stationary regime, while in the mentioned experiments,
this was the position of the second yielding point.
It is unlikely that the full angle-resolved ITT-MCT will recover the
two-step yielding, and first numerical results confirm this
\cite{AmannJOR}.
Apart from appealing to the known aging- and hopping-induced relaxation
phenomena that become relevant in the attractive glass,
a possible resolution of this discrepancy may lie in the fact that the
second yielding can be attributed to the breaking up of larger clusters of
particles, reminiscent of the heterogeneous nature of the attractive glass.
Such aspects are not captured in the present ITT-MCT.

This issue notwithstanding, one may identify the yield strain obtained by
ITT-MCT with the first yielding point observed in experiment. The
predictions of the theory regarding its dependence on attraction strength
and range can then in principle be tested.

\section{Conclusions}\label{conclusion}

We have discussed aspects of the steady-state nonlinear rheology and yielding
of dense colloidal
suspensions of hard-core particles with a short-ranged attraction,
within ITT-MCT and an isotropic approximation to its correlation functions.
Our emphasis was on the transition from repulsive to attractive glasses
and the manifestation of attraction-induced reentrant melting and
glass--glass transitions in nonlinear rheology. The dynamic yield stress,
defined as the $\dot\gamma\to0$ limit of the steady-state flow curves,
was found to increase strongly with the attraction strength, while it
is largely independent on the attraction range. Similar conclusions hold
for the maximum stress observed in the stress--strain curves measured
in startup flow; a quantity that is often also referred to as a yield
stress. The corresponding yield strain scales with the attraction range
at strong attractions, and reflects the interplay of weakening and
stabilizing of the glass at low attraction strengths.
The first normal-stress difference was found to be less sensitive to
attraction than the shear stress.

While the phenomena of double-yielding in startup flow that were reported
in experiment are not explained by the current theory, this failure may
be connected to additional cluster formation at very strong attractions.
It will be interesting to see whether the ITT-MCT predictions can
be found for systems with weak attraction strengths.

In applying ITT-MCT to attractive glasses one should recall that the
theory, at present, presupposes the existence of a homogeneous flow field.
This issue also arises in repulsive glasses, where shear banding occurs
in certain parameter regimes \cite{Besseling2010}.
Attractive systems, however, may be more susceptible to shear banding
\cite{Becu2006}.

It will be intriguing to see whether a combination of ITT-MCT with the
cluster-MCT proposed by \textcite{Kroy2004} can explain the
second yield step reported for attractive glasses. The cluster-MCT approach
combines (quiescent) MCT with the notion of an RG-like flow in parameter space,
in order to account for the slow aggregation of particles.

\begin{acknowledgments}
We thank M.~Fuchs for discussions.
M.~P.\ thanks for funding from DLR-DAAD fellowship No.~141.
Th.~V.\ thanks for funding from the Helmholtz Gemeinschaft (HGF VH-NG 406).
\end{acknowledgments}

\bibliography{lit}
\bibliographystyle{apsrev4-1}

\end{document}